\documentclass{aa}
\usepackage{graphicx,amsmath}
\usepackage{txfonts}

\usepackage{natbib} 
\bibpunct{(}{)}{;}{a}{}{,}

\begin{document}
\title{Spectroscopy at the solar limb: I.~Average off-limb profiles and Doppler shifts of \ion{Ca}{ii}\,H}

   \author{C.\,Beck\inst{1,2} \and R.\,Rezaei\inst{3}}
        
   \titlerunning{Spectroscopy at the solar limb: I.~Average profiles and Doppler shifts of \ion{Ca}{ii}\,H}
  \authorrunning{C.Beck \& R.Rezaei}  
   \institute{Instituto de Astrof\'{\i}sica de Canarias, Spain     
     \and Departamento de Astrof{\'i}sica, Universidad de La Laguna, Spain
     \and Kiepenheuer-Institut f\"ur Sonnenphysik, Germany
    }
\date{Received xxx; accepted xxx}
\keywords{Sun: chromosphere -- Techniques: spectroscopic -- Methods: data analysis}
\abstract{}{We present constraints on the thermodynamical structure of the chromosphere from ground-based observations of the \ion{Ca}{ii}\,H line profile near and off the solar limb.}{We obtained a slit-spectrograph data set of the \ion{Ca}{ii}\,H line with a high signal-to-noise ratio in a field of view extending 20$^{\prime\prime}$ across the limb. We analyzed the spectra for the characteristic properties of average and individual off-limb spectra. We used various tracers of the Doppler shifts, such as the location of the absorption core, the ratio of the two emission peaks H$_{\rm 2V}$ and H$_{\rm 2R}$, and intensity images at a fixed wavelength.}{The average off-limb profiles show a smooth variation with increasing limb distance. The line width increases up to a height of about 2 Mm above the limb. The profile shape is fairly symmetric with nearly identical H$_{\rm 2V}$ and H$_{\rm 2R}$ intensities; at a height of 5 Mm, it changes into a single Gaussian without emission peaks. We find that all off-limb spectra show large Doppler shifts that fluctuate on the smallest resolved spatial scales. The variation is more prominent in cuts parallel to the solar limb than on those perpendicular to it. As far as individual structures can be unequivocally identified at our spatial resolution, we find a specific relation between intensity enhancements and Doppler shifts: elongated brightenings are often flanked all along their extension by velocities in opposite directions.}{The average off-limb spectra of \ion{Ca}{ii}\,H present a good opportunity to test static chromospheric atmosphere models because they lack the photospheric contribution that is present in disk-center spectra. We suggest that the observed relation between intensity enhancements and Doppler shifts could be caused by waves propagating along the surfaces of flux tubes: an intrinsic twist of the flux tubes or a wave propagation inclined to the tube axis would cause a helical shape of the Doppler excursion, visible as opposite velocity at the sides of the flux tube. Spectroscopic data allow one to distinguish this from a sausage-mode oscillation where the maximum Doppler shift and the tube axis would coincide.}
\maketitle
\section{Introduction}
The solar chromosphere is a very variable atmospheric layer governed by dynamical processes. Part of its organization is provided by the photospheric magnetic fields and their extension into the upper solar atmosphere, but the main characteristic of the chromosphere is its dynamical evolution. The chromospheric temperature is higher than that of the underlying photosphere, but because the gas is transparent owing to its low density, energy sources other than radiation have to maintain the pure existence of the chromosphere \citep[see, e.g.,][and references therein]{rutten2010}. Different methods have been suggested to provide the necessary energy, ranging from a purely mechanical transfer of energy by propagating waves to various processes related to the magnetic field such as reconnection \citep[][]{biermann_48,nar_ulm_96,kalkofen1996,carlsson+stein1997,fossum+carlsson2005,navarro2005}. 

Owing to its low gas density, the assumption of local thermal equilibrium (LTE) is not justified in the chromosphere. Spectral lines like H$\alpha$, \ion{Ca}{ii} H, the \ion{Ca}{ii} IR triplet, or \ion{He}{i} at 1083\,nm contain the information on the thermodynamical and magnetic properties of the chromosphere \citep[e.g.,][]{bala+etal2004,socasnavarro+elmore2005,pietarila+etal2007,rezaei+etal2007,beck+etal2008,cauzzi+etal2008,centeno+etal2010}, but the non-LTE conditions of the line formation complicates the analysis of the spectra. Because there are relatively few approaches to directly derive the desired information from individual spectra, the investigation of the chromosphere is often made with average spectra of some sort. This allowed static atmospheric models corresponding to a temporally and spatially averaged chromosphere to be developed \citep[][]{vernazza+etal1981,fontenla_etal_06}, but their relevance for individual spectra is under debate \citep{fontenla_etal_07}. 

For these studies, usually spectra observed on the center of the solar disk were used. Because most of the chromospheric spectral lines have a significant contribution from photospheric layers in their line wings, disk-center spectra can be contaminated with photospheric information as well. Observations near and off the solar limb remove the photospheric contribution because the line of sight (LOS) for observations above the solar limb is not passing through photospheric layers anymore.

We aim to improve the observational constraints on the solar chromosphere by investigating medium-resolution \ion{Ca}{ii}\,H spectra near and above the limb. The data were obtained in 2009 at the German Vacuum Tower Telescope \citep[VTT,][]{vtt}, Iza{\~n}a. Thanks to the Kiepenheuer Institute adaptive optics system \citep[KAOS;][]{vdluehe+etal2003}, they are intermediate between the high-resolution imaging data in \ion{Ca}{ii}\,H from the HINODE satellite \citep{kosugi+etal2007} and the spectroscopic data obtained before the advent of adaptive optics (AO) systems \citep{zirker1962,beckers68,zirker1968}, providing a full line-spectrum with high spectral resolution. The observations are described in Sect.~\ref{sect_obs}. The data reduction is discussed in Sect.~\ref{sect_datared}, whereas Sect.~\ref{sect_results} displays our observational findings. Section \ref{sect_disc} summarizes our results, which are discussed in Sect.~\ref{sect_concl}. Section \ref{sect_concll} presents our conclusions.
\section{Observations\label{sect_obs}}
During an observation campaign for a study of the center-to-limb variation
(CLV) of \ion{Ca}{ii}\,H spectra in August 2009, we obtained some data sets at
the very limb as well. We selected the one with the best spatial resolution for 
the present study, taken on 25 Aug 2009 between 08:43 and 09:15\,UT. 
The spectra were taken with the POlarimetric LIttrow Spectrograph \citep[POLIS,][]{beck+etal2005b} at the VTT, a slit-spectrograph spectropolarimeter with two channels at 630\thinspace nm (``red channel'') and in \ion{Ca}{ii}\,H at 396.85\thinspace nm (``blue channel''). The observed wavelength range of the red channel covered a range slightly more toward the blue than the standard configuration, containing the \ion{Fe}{i} line at 630.15\thinspace nm and the \ion{O}{i} line at 630.06\thinspace nm instead of the \ion{Fe}{i} line at 630.25\thinspace nm. Since the removal of the polarizing beamsplitter in the blue channel in 2005 to increase the low light level in the near-UV, the Ca channel provides spectroscopy only. We used an integration time of about 13 sec per scan step. The spectral sampling in the Ca channel was 1.95\,pm per pixel at a nominal resolution of the
  spectrograph of $\Delta\lambda/\lambda=\!220,\!000$.  The slit width corresponded to 0\farcs5, and the slit was moved in 134 steps of 0\farcs3 across the solar image, yielding a field of view (FOV) of about 40$^{\prime\prime} \times 70^{\prime\prime}$. The spatial sampling along the slit corresponded to about 0\farcs3. The center of the FOV was located on the solar disk at $(+37^{\prime\prime},+920^{\prime\prime})$, giving a heliocentric angle of 74 degree. The slit was inclined by  30 deg to the solar limb and protruded at maximum about 20$^{\prime\prime}$ over the limb in the Ca channel. The KAOS system was locked on a bright facula near the center of the FOV. Because at the VTT the scanning system of either the main spectrograph or POLIS is independent of the AO, the AO could stay locked on the same point of the solar surface except for one failure (x$\sim$\,23\arcsec, marked in Fig.~\ref{fig2b} below) where the AO lock point jumped by about 2$^{\prime\prime}$ to another near-by facula.

During the whole CLV campaign, we intermittently took data sets with the same 
integration time on the solar disk center in between the CLV observations. These disk center data sets allowed us to derive an intensity normalization curve with which we were able to normalize the intensity of each observation in (arbitrary) detector counts to the disk-center intensity at the time of the observations \citep[cf., e.g.,][]{zirker1968}. In addition to the regular observations, we also carried out several types of measurements to determine the stray-light level in the spectra. Here we will only discuss an observation where half of the FOV was blocked by a metal plate in the telescope focal plane up front of the AO. The data was used to derived an estimate of the off-limb stray light and the spatial point spread function of POLIS. The measurements for determining other stray-light contributions and their evaluation are discussed in more detail in another publication \citep[][submitted to A\&A]{beck+rezaei2010a}.
\begin{figure}
\centerline{\resizebox{8.8cm}{!}{\includegraphics{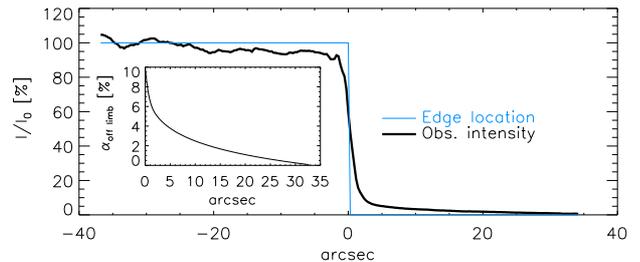}}}
\caption{Intensity variation along the slit near 396.4\thinspace nm for a partly blocked FOV. {\em Thick black}: 
observed intensity variation. {\em Blue}: location of the sharp edge. The inset shows the derived stray-light contribution $\alpha$ as function of the limb distance.\label{fig_1a}}
\end{figure}

We estimate the spatial resolution of the data to be about 1$^{\prime\prime}$\,--\,1.5$^{\prime\prime}$, with a residual  uncorrected image motion of less than 0\farcs5 (see Appendix \ref{spat_res_sect}).
\section{Data reduction\label{sect_datared}}
The data of both POLIS channels were reduced by the approaches described in \citet{beck+etal2005b,beck+etal2005a}. This includes the correction for dark current and the correction for flat-field imperfections, and in the case of the Ca channel a correction for the order-selecting pre-filter in front of the camera.  To improve the data quality, we added some more reduction steps before we evaluated the spectra, however. \paragraph{Stray light correction} Because in the end we aim to arrive at a conversion from observed intensities in detector counts into the absolute energy of the radiation, the data needed to be corrected for stray-light contributions. In this context, we modeled the spectrum $I(\lambda)$ of a certain pixel $(x,y)$ in the FOV as 
\begin{equation}
I_{\rm obs} (\lambda,x,y) = I_{\rm true} (\lambda,x,y) + \alpha \cdot I_{\rm  local}(\lambda) + \beta \cdot <I_{\rm local}>   \label{eq_stray_simp}\;,
\end{equation}
where $I_{\rm  local}(\lambda)$ is some average profile of the local FOV, and
$<I_{\rm local}>$ an average intensity value without wavelength
dependence. From the analysis of the stray-light measurements mentioned in the
previous section, we derived typical values of $\beta$ = 5\% and $\alpha$ =
10\% as lower limit across the full FOV for the contributions of the parasitic\footnote{Spectrally un-dispersed light that still reaches the CCD cameras by scattering inside the instrument without passing through the grating.} and local stray light, respectively. The observed spectra located on the solar disk were thus corrected by subtracting 5\% of the intensity near 396.4\thinspace nm in the average profile from the whole spectrum, and then 10\% of the average spectrum $I_{\rm  local}(\lambda)$ was subtracted from each profile. 
\paragraph{Off-limb correction} It became clear from the observed spectra where the slit crossed the limb, or from the measurement with the half-blocked FOV that the stray-light contribution $\alpha$ depends on the distance to the limb (or the blocking edge). To take this effect into account, we scaled $\alpha$ down with increasing limb distance by the spatial intensity variation that was observed with the half-blocked FOV {({\em black line} in Fig.~\ref{fig_1a}). We cut the observed intensity curve at the approximate location of the edge and normalized the first point to 10\% (value of $\alpha$ on the disk; see the inset). The reduced $\alpha$ was then used for all off-limb spectra as a function of their distance $d$ to the ``white-light'' limb seen at about 396.4\thinspace nm. 

It turned out that subtraction of $\alpha(d) \cdot I_{\rm local}$ with $\alpha(d=0^{\prime\prime}) = 10\,\%$ did not yield a good correction beyond the limb because the off-disk stray light was not fully compensated. We first tried to use an average disk-center profile with its higher intensity level instead of $I_{\rm local}$, but this over-compensated the stray light and left strong residuals of the line blends in the Ca wing because they were located at slightly different wavelengths in the average disk center profile. We therefore decided to multiply the intensity of the average profile $I_{\rm local}$ with a scaling coefficient, and by trial-and-error obtained that an increase by a factor of 2 gave a satisfactory correction off the disk. We first attributed this to the fact that the scanned FOV is significantly smaller than the full FOV that passes through the field stop of POLIS, thus there is also an area with higher light level than near the limb. An in-depth study of the stray-light level \citep[][submitted to A\&A]{beck+rezaei2010a} using a semi-empirical spatial point spread function (PSF) for
POLIS later revealed that the stray light from the close surroundings of a
given spatial point in the quiet Sun is indeed closer to $\alpha=$\,20\,\% than the $\alpha=$\,10\,\% used that was derived from profiles in the umbra of sunspots. The stray-light correction uses $\alpha\,I_{\rm local}$ in
Eq.~(\ref{eq_stray_simp}), where the factor of 2 can therefore be attributed to
either $\alpha$ or $I_{\rm local}$.  If one attributes it to $\alpha$ as
suggested by the PSF, the effective stray-light correction consisted of a correction for 20\,\% stray light on-disk using the average profile of the observation $I_{\rm local}$, and for a stray-light coefficient $\alpha(d)$  beyond the limb corresponding to {\em twice} the values shown in the inset of Fig.~\ref{fig_1a}. 

The observation of the knife-edge function with the half-blocked focal plane only covers the optics downstream of the telescope focal plane. Thus neither the PSF of the telescope nor fluctuations of the scattering in the Earth's atmosphere are taken into account in the stray-light correction. Fluctuations of the scattering presumably were minor during the observations (see Fig.~\ref{limb_pos1}). The stray-light correction is smooth across the limb location because $\alpha(d=0^{\prime\prime})$ was normalized to the on-disk value, which still holds regardless if $\alpha$ or $I_{\rm local}$ are scaled up, and because the doubled stray-light amount $\alpha\,I_{\rm local}$ with $\alpha=20\,\%$ was subtracted on the disk.

This approach for the off-limb correction yielded a residual intensity of
below about 0.2\% of $I_c$ for all off-limb spectra and wavelengths outside the Ca core without significant residuals of the line blends in the Ca line wing (see Fig.~\ref{fig5} below). The procedure corrects for the contamination with scattered light from the solar disk inside the local FOV in the direction
perpendicular to the limb, but does not account for a possible contamination
with stray light from the disk center \citep[see,
e.g.,][]{zwaan1965,mattig1971,martinezpillet+etal1990}. Using an average disk center profile for the stray-light correction worked worse for our data, partly because the wavelength locations of the spectral lines in the wing of \ion{Ca}{ii} H in the disk center profile did not match to the locations in the stray light beyond the limb. The stray-light contamination close to the limb therefore seems to come mainly from inside the local FOV, as also found by \citet{mattig1983}. The light level in (pseudo)continuum windows of the spectra beyond the limb also was only about 2\% of $I_c$ {\em before} the correction.
\paragraph{Intensity normalization} After the correction for the stray-light contamination, the spectra were normalized to absolute units by two coefficients. The first coefficient was derived from the series of disk center measurements to obtain the detector count value in the line wing at disk center at the moment of the observation. After division with this value of about 3000 counts, the spectra are normalized to unity in the line wing on the disk center. To facilitate a conversion to absolute energy units, we then determined the normalization coefficient of an average disk-center spectrum taken on the same day to the FTS atlas \citep{kurucz+etal1984} at the same wavelength in the line wing ($I({\rm FTS,396.4\thinspace nm})\sim$\,0.357). Because we also have a Ca spectrum on disk center in absolute energy units from an NLTE calculation available \citep{rezaei+etal2008}, we were then able to convert all relative intensity values to absolute energy units. Figure \ref{fig1} shows the average spectrum of the observed FOV after all corrections and the normalization; the intensity scale relative to the continuum intensity is given {\em at the left}, the actual observed detector count values are provided by the scale {\em at the right}. Even with the 13-sec integration, the low light level in the Ca core led to only a few hundred counts.
\begin{figure}
$ $\\
\centerline{\resizebox{8.8cm}{!}{\includegraphics{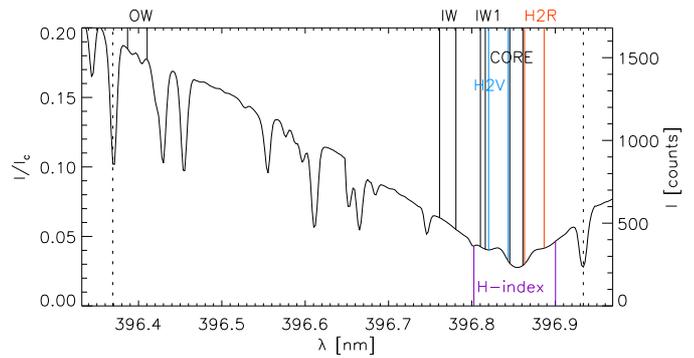}}}
\caption{Average Ca spectrum of the observed FOV (on disk). The locations of specific spectral windows are marked by {\em vertical solid lines}. The {\em vertical dashed lines} denote two blends whose line-core velocity was additionally measured. The {\em left} scale gives the intensity relative to the disk-center continuum, the {\em right} scale the detector count value before the normalization.\label{fig1}}
\end{figure}

We marked several wavelength windows in the spectrum that will be used below, with increasing wavelength outer wing (OW), inner wing (IW) and inner wing 1 (IW1), H$_{\rm 2V}$ (396.832$\pm$0.012\,nm), line core (CORE, 396.853$\pm$0.008\,nm), and H$_{\rm 2R}$ (396.875$\pm$0.012\,nm) \citep[cf.~also the definitions in][]{cram+dame1983,lites+etal1993}. The commonly used H-index (an 1 {\AA} wide band around the line core) is marked as well. We did not calculate the integral over the spectral window for all these quantities but the average value relative to $I_c$ instead, which allows an easier comparison with individual spectra or the intensity at some fixed wavelength.
\section{Data analysis and results\label{sect_results}}
The \ion{Ca}{ii}\,H line spectra are usually complex-shaped in the interesting
wavelength range near the Ca line core, and no direct analysis tools exist for
a derivation of solar atmospheric properties from the spectra, mainly because of the NLTE conditions under which the line core forms. The Ca spectra
near and off the limb are even more difficult to analyze than Ca spectra on
the disk. This is caused by the effect of the integration along the LOS. On the disk, one can assume that a LOS traverses a very limited number
of different structures because of the steep increase in density and thus
opacity along the LOS, whereas a LOS near the limb can pierce many more
structures in an atmosphere of nearly constant density along the LOS. It
turned out, however, that most analysis methods employed on disk spectra
\citep[cf.][]{rezaei+etal2007,rezaei+etal2008,beck+etal2008,beck+etal2009} can
be used on the limb  spectra as well. 
\begin{figure}
\centerline{\resizebox{8.8cm}{!}{\includegraphics{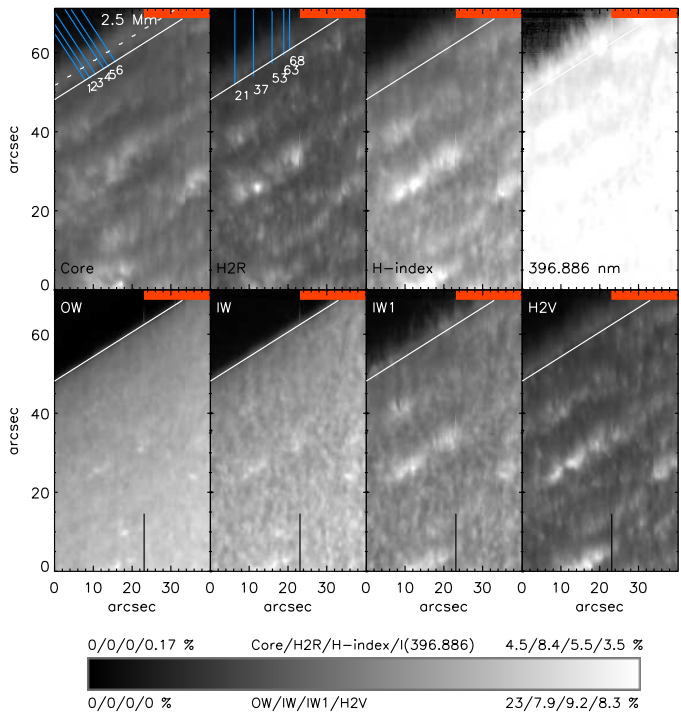}}}
\caption{Overview of the FOV in different spectral windows, denoted at the
  {\em upper/lower left} inside each image. The {\em inclined white line} denotes the limb, the {\em black vertical line} at the bottom at $x\sim 23^{\prime\prime}$ a jump of the AO. For this reason, the {\em red shaded} area in the upper right corner of each image could not be used after alignment to the rest of the FOV. The {\em blue lines} in the {\em first two columns} of the {\em top row} denote cuts through the FOV used later on; the {\em dashed white line} in the line-core map marks a limb distance of 2.5 Mm. The {\em fourth column} in the {\em top row} is taken at a single fixed wavelength and thresholded to enhance the visibility of off-limb structures.\label{fig2b}}
\end{figure}
\begin{figure*}
$ $\\$ $\\
\centerline{\resizebox{17.6cm}{!}{\includegraphics{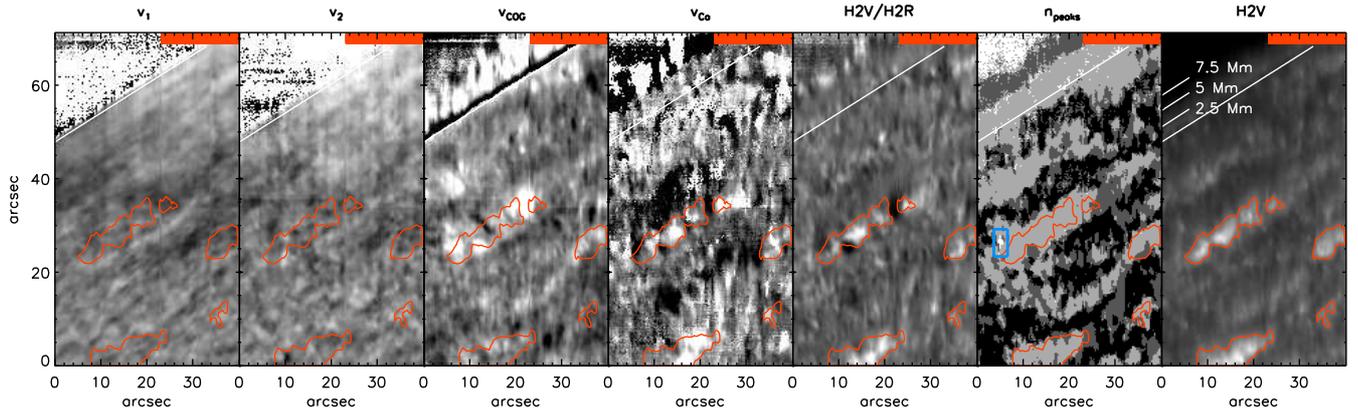}}}
\caption{{\em Left to right}: line-core velocity of 396.369\,nm, same for  396.93\,nm,  COG of Ca core region, velocity of Ca core, ratio H$_{\rm 2V}$/H$_{\rm 2R}$, number of reversals from 0 ({\em black}), 1 ({\em dark gray}), 2 ({\em light gray}) to 3 or more ({\em white}), H$_{\rm 2V}$ intensity. {\em White/black} in all velocity maps corresponds to motions away from/toward the observer. The {\em red contour lines} denote locations with increased emission. The {\em blue rectangle} in the peak map outlines a region of multi-lobed profiles. The short bars parallel to the limb in the  H$_{\rm{2V}}$ map denote some heights above the limb. \label{fig3}}
\end{figure*}
\subsection{Full FOV}
\paragraph{FOV in different wavelength ranges} Figure \ref{fig2b} shows how
the FOV appears in the various wavelength ranges marked in Fig.~\ref{fig1}. We
defined the location of the limb in ``pseudo-white-light'' as the place where the intensity drops below 5\% in the OW map. The (isolated) bright faculae in the
OW map are all related to a significant polarization signal in the 630\thinspace nm channel (not shown here). From OW to IW, the on-disk features increase in contrast, and generally have the trend to increase in size as well. The intensity seen beyond the limb is increasing with decreasing distance to the very line core. The map of IW1 is the first one where the structuring changes from roundish shapes to elongated features. Several streaks with high intensity can be seen above the limb that were absent before. The orientation of the streaks is closer to being parallel to the slit than to being perpendicular to the limb in many cases. This could either reflect the true orientation of the underlying solar structures or be somehow caused by the way of data acquisition with a slit-spectrograph system, but the spatial resolution achieved does not allow one a definite conclusion (compare to, e.g., \citet{pasachoff+etal09}, Fig.~13, or \citet{shoji+etal2010}, Fig.~2). The fine-structure above the limb can also be clearly seen in the intensity map taken at a fixed wavelength ({\em fourth column} in the {\em top row}) whose display range was thresholded to emphasize the off-limb intensities. In the latter, the off-limb fine-structure actually is more prominent than in the maps of ``pure'' chromospheric quantities like H$_{\rm 2V}$, line core, and  H$_{\rm 2R}$. The emission extends furthest above the limb in the line-core map, as is expected, but in general is more uniform there than for H$_{\rm 2V}$ and  H$_{\rm 2R}$ \citep[similar to the Fig.~7 of][]{rutten2007}. If one compares the H$_{\rm 2V}$ and H$_{\rm 2R}$ maps with the wavelength-integrated H-index ({\em third column} in {\em top row}), one finds several occasions where the bright emission in the H-index actually only comes from one of the emission peaks, e.g., at the lower middle near $(x,y)= (18^{\prime\prime},2^{\prime\prime})$ or near $(x,y)=(11^{\prime\prime},23^{\prime\prime})$. The H-index does not allow one to distinguish between the two cases.
\paragraph{LOS velocity proxies} We determined the LOS velocities for two of the
photospheric blends ({\em dotted lines} in Fig.~\ref{fig1}) that correspond to
the lowest (v$_1$, \ion{Cr}{i} at 396.369\thinspace nm) and highest (v$_2$,
\ion{Fe}{i} at 396.93\thinspace nm) forming lines inside the observed spectral
range \citep{beck+etal2009}. For the Ca core, we also determined the location
of the minimum intensity in each profile. This method, however, fails from
some distance above the limb outward because the Ca core shows a single
emission peak there. For that region, the ratio between  H$_{\rm 2V}$ and
H$_{\rm 2R}$ emission peaks has to be used as proxy for the velocity{;
  \citet{rezaei+etal2007} discussed the tight relation between the location of
  the absorption core and the H$_{\rm 2V}$/H$_{\rm 2R}$ ratio (their
  Fig.~6)}. The H$_{\rm 2V}$/H$_{\rm 2R}$ ratio still indicates a Doppler shift when the profile reverts to a pure emission pattern, but with the opposite velocity sign. We also calculated the center-of-gravity (COG) of the intensity
in an 1.5 {\AA}  range around the Ca line core, but this method is only reliable above the limb where the Ca profiles show a single emission peak. The zero point of the velocity was defined by the average value in the lower half of the FOV. Figure \ref{fig3} shows the resulting 2-D maps for the velocity proxies. 
The patterns of photospheric and chromospheric velocities are fairly
  unrelated; the chromospheric velocity pattern also shows no relation to the
  photospheric OW map \citep[see also][]{beck+etal2009}. The two maps of the
line-core position (v$_{\rm Ca}$, {\em fourth column}) and the ratio of
H$_{\rm 2V}$/H$_{\rm 2R}$ ({\em fifth column}) match in the spatial  {patterns
  on the disk, e.g., inside and close to the {\em red contours}}, even if the
amplitude of changes is higher for v$_{\rm Ca}$. This supports the use of the ratio of H$_{\rm 2V}$ and H$_{\rm 2R}$ for the spectra above the limb without any clear absorption core.
\paragraph{Profile shape analysis} For the analysis of the profile shape, we used the same method as in \citet{rezaei+etal2008}, the determination of the number of intensity reversals. The reversals are defined as local
maxima of the intensity spectrum in a wavelength range around the line
core. The resulting map is displayed in the {\em sixth column} of Fig.~\ref{fig3}. 

In comparison to observations near disk center, the area fraction of double-peaked profiles is slightly enhanced near the limb ($> 50\%$). This could be an effect of the strongly inclined LOS: upward propagating waves have a smaller LOS component at the location of the FOV, therefore the central absorption core is located at the rest wavelength between the two emission peaks instead of suppressing  H$_{\rm 2R}$ as happens on disk center. Several coherent larger-scale areas can be found where profiles without reversals exist \citep[{\em black shading}, cf.][]{rezaei+etal2008}, located between the ``magnetic'' regions as indicated by the increased emission and the polarization signal. From the very limb to about 5 Mm height, the profiles all have at least two reversals ({\em light gray shading}). At about 5 Mm there is a rather abrupt transition to single-lobed profiles ({\em dark-gray shading}); the height of the transition shows little variation with the spatial position along the limb. We visually inspected the profiles with more than two reversals ({\em white}). Near and above the limb they turned out to be artifacts created by either the spectral line of \ion{Fe}{i} at 396.93\thinspace nm or the noise level, but in one region still on the disk ({\em blue rectangle} in the {\em sixth column} of Fig.~\ref{fig3} at $(x,y) =
(5^{\prime\prime},25^{\prime\prime})$\,) they are real. Figure \ref{multi_lobe}
shows one such profile with several reversals near the line core. It would be interesting to see if numerical simulations like those described by \citet{wedemeyer+etal2004} or \citet{martinez+etal2009} yield such profiles in a spectral synthesis \citep[for an initial attempt, see][]{haberreiter+etal2010}.
\begin{figure}
\centerline{\resizebox{8.8cm}{!}{\includegraphics{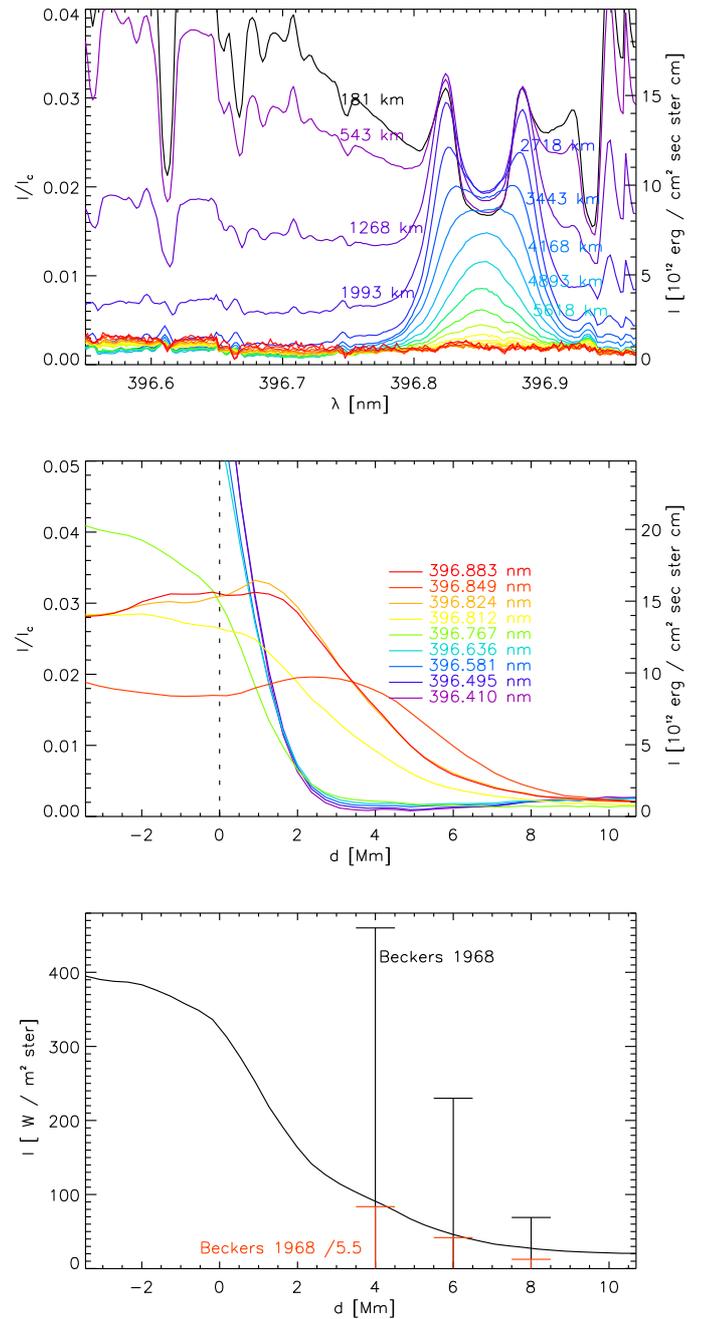}}}
\caption{ {\em Top panel}: average off-limb spectra of \ion{Ca}{ii}\,H with increasing limb distance $d$. {\em Middle panel}: intensity variation with limb distance for the wavelengths marked in Fig.~\ref{fig42}. The limb is denoted by a {\em vertical black dashed line}. The {\em right} scale uses the same units as in \citet{zirker1962}. {\em Bottom}: the energy of the radiation integrated over the line-core region. The {\em black vertical bars} denote the energy as given by BE68, and after division by a factor of 5.5, respectively ({\em red}).\label{fig5}}
\end{figure}
\subsection{Average off-limb spectra}
Figure \ref{fig5} shows the evolution of the spectral shape of \ion{Ca}{ii}\,H
with increasing distance to the limb, $d$. The spectra were averaged over
height bins of 362 km ($\equiv$ 0\farcs5) and across the FOV; except for the two profiles closest to the limb, only every second profile is plotted in the {\em uppermost panel}. The profiles are labeled with the corresponding height. Close to the limb, the average spectra are double-peaked with only a small asymmetry
between H$_{\rm 2V}$ and H$_{\rm 2R}$. The line-core intensity slightly increases when going outward, whereas the emission peaks decrease. At a height of about 4-5 Mm above the limb this leads to profiles without a central absorption, showing a broad plateau instead. The central emission decreases slower than that near the wavelengths of H$_{\rm 2V}$ and H$_{\rm 2R}$ and a single-peaked broad Gaussian remains above about 5 Mm. Extrapolating the intensities measured by \citet{dunn+etal1968} during an eclipse, \citet{linsky+avrett1970} suggested that at a height of about 5-6 Mm above the limb the ratio of the \ion{Ca}{ii}\,H and K lines should approach a value of 2, implying that the lines become optically thin. Then the central absorption disappears, leaving a single emission core, which agrees with our observed spectra. The profiles compare well to those given by \citet[][ZI62, his Fig.~3]{zirker1962}, generally having a slightly lower intensity. In a recent publication, \citet{judge+carlsson2010} provided synthetic Ca spectra for a simplified model of spicules near and off the limb. The spectra shown in their Fig.~3 seem to match the general properties found in our observations well, showing a similar transition from double-peaked emission to a single-peak emission of decreasing width with height.

The \ion{Fe}{ii} line near 396.94\thinspace nm reverts to emission, as do some of the other blends, but none as strong as this one \citep[see][]{engvold+halvorsen1973,rutten+stencel1980}. This line is consequently interesting for detailed quantitative analyses as well \citep{lites1974,watanabe+steenbock1986}, to avoid solely using the \ion{Ca}{ii}\,H line core, even if \citet{cram+etal1980} found a photospheric origin for the emission in this line. 
 
All wavelengths that show emission have a (more or less) pronounced maximum of intensity whose height above the limb increases the closer the wavelength is to the line core ({\em middle panel}). This is similar to the findings of \citet{white1963} for the \ion{He}{i} D$_3$ line or for the \ion{He}{i} line at 1083\thinspace nm \citep{schmidt+etal1994,penn+jones1996,sanchezandrade+etal2007}. The maximum intensity occurs at layers higher than found using intensity contribution functions for these wavelengths \citep[e.g.,][]{rezaei+etal2008,beck+etal2009}. This is caused by the stretched optical depth scale in a LOS nearly parallel to the solar surface. 

We determined the energy contained in the radiation emitted by the line-core
region by integrating the spectra over the wavelength range from
396.72\thinspace nm to 396.96\thinspace nm ({\em bottom panel}). The total
energy radiated by the \ion{Ca}{ii}\,H line smoothly decreases from about 400 Wm$^{-2}$ster$^{-1}$ near the limb on the disk to about 100
Wm$^{-2}$ster$^{-1}$ at a height of 4 Mm \citep[cf.][]{judge+carlsson2010}. The values provided by BE68 for heights of 4 and 6 Mm in his Table XI match our values when divided by a factor of 5.5 ({\em red vertical bars}). This is presumably because of his scaling of the total intensity to an ad-hoc assumed spicule size, whereas we use spectra of single pixels corresponding to a 0\farcs5\,$\times$\,0\farcs3 area. 

Figure \ref{fig42} shows the full set of average off-limb spectra. In the {\em
  top panel} we normalized  each row of the spectrum separately to its
maximum intensity between 396.763\thinspace nm and 396.920\thinspace nm to
highlight the shape. We then used three different methods to estimate the line
width of the spectra. For spectra up to about 4 Mm, two distinct emission
peaks are seen whose position can be determined ({\em blue dots}). After the
emission peaks have merged to a single broad peak, we determined the location
where the intensity drops to 50\% of the maximum intensity near the line core
and fitted a Gaussian to the core region of the spectra as well to derive the
full width at half maximum (FWHM, {\em black dots}). The latter two methods
gave similar results, but cannot be employed below a limb distance of about 2 Mm because the increasing intensity of the line wings makes it impossible to
define where the contribution of the line core ends. For heights below 2 Mm
one could assume that the FWHM will follow the behavior of the emission peaks,
which would yield the {\em black dash-dotted line} in the {\em lower panel} of Fig.~\ref{fig42}. The FWHM of the emission signal near the line core, including the H$_{\rm 2V}$ and H$_{\rm 2R}$ emission peaks, increases slightly from about 90 pm at the limb to about 97 pm at 2 Mm height, then decreases by a factor of 2 until about 5 Mm \citep[cf.~for instance][his Fig.~9]{makita2003}. The subsequent increase of the width for layers above 5 Mm is presumably not real because the intensity reaches the noise level. Our values agree to first order with the line widths provided in ZI62 (83\,pm@5.1\,Mm and 68\,pm@6.4\,Mm) or in BE68. The latter defined the width $w$ as FWHM/$\lambda$ (see the scale {\em at right}) and obtained $w=1.4$ at 5 Mm (his Table XI). 
\begin{figure}
\centerline{\resizebox{7.cm}{!}{\includegraphics{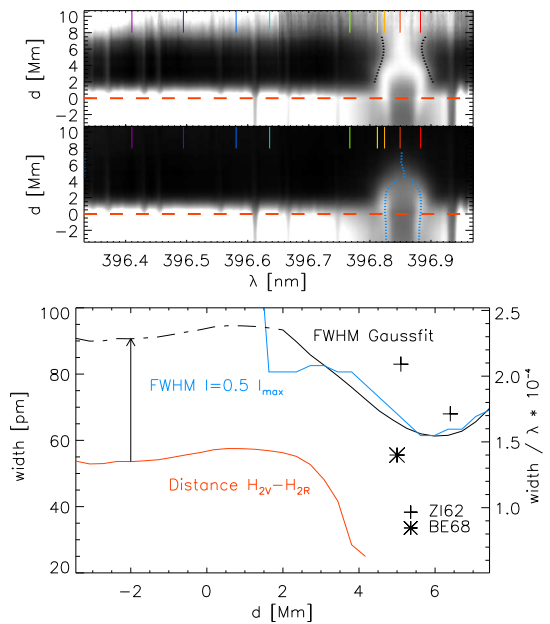}}}
\caption{ {\em Top}: the full average off-limb spectrum {\em normalized row by row}; the wavelengths used in the middle panel of Fig.~\ref{fig5} are marked by short {\em vertical lines}  at the top.  {\em Middle panel}: the same spectrum without additional normalization. The {\em blue dots} indicate the locations of the emission peaks, the {\em black dots} the FWHM. The {\em red horizontal dashed lines} denote the location of the limb. {\em Bottom panel}: FWHM from a fit of a Gaussian ({\em black}), from the location of 50\% of the maximum intensity ({\em blue}), and the distance between the emission peaks ({\em red}). {\em Crosses} and {\em asterisks} mark the values given by ZI62 and BE68, respectively. \label{fig42}} 
\end{figure}

The observations beyond the limb offer the possibility to define the FWHM of
the \ion{Ca}{ii}\,H line for a Sun as a star spectrum. The values near a
height of 2 Mm in Fig.~\ref{fig42} suggest that the FWHM is larger than the
peak separation by a factor of about 1.7 \citep[see also][their Figs.~6 and 8
yield a factor of about 1.4]{sivaraman+etal1987}. The distance between the
H$_{\rm 2V}$ emission peak and the line core in the FTS atlas spectrum is
about 15.6\thinspace pm. Because the position of H$_{\rm 2R}$ cannot be reliably
determined in the profile, we assume that the peak separation should be twice
as large. This finally gives a FWHM at disk center of 53\thinspace pm,
corresponding to 40 kms$^{-1}$ (logarithmic value of
1.6). \citet{wilson+bappu1957} give a range of 33-39 kms$^{-1}$ for Ca spectra
in plage where the emission core is more pronounced and can be separated from
the wing contribution \citep[see
also][]{teplitskaja+efendieva1973,jebsen+mitchell1978,oranje1983,sivaraman+etal1996}. The prediction of \citet{wilson+bappu1957} for the Sun as a star line width of \ion{Ca}{ii}\,H or K is 1.56, derived from their curve of what later was
termed the Wilson-Bappu effect \citep[cf.~also][]{stencel2009}.
\begin{figure}
\centerline{\resizebox{8.8cm}{!}{\includegraphics{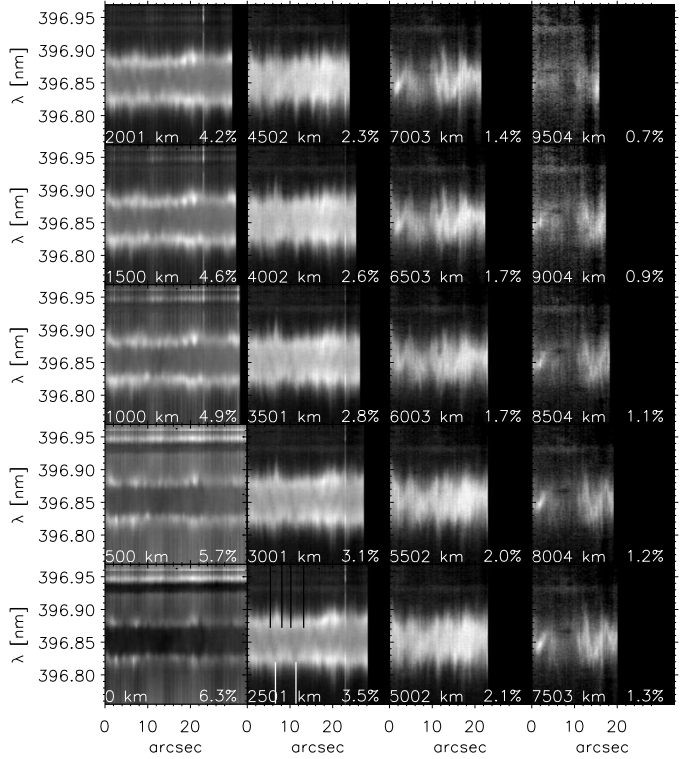}}}
\caption{Spectra on cuts parallel to the limb. The height of the cut is denoted in the {\em lower left}, the maximum intensity 
at {\em lower right}; the x-axis gives the position along the limb, the y-axis the wavelength. For the spectra at 2.5 Mm, the 
locations of Doppler excursions of the absorption core to the blue/red are denoted by {\em white/black vertical bars}.\label{fig6}}
\end{figure}
\subsection{Individual off-limb spectra}
We discuss the spatially resolved spectra on cuts parallel to the limb, along the slit, and perpendicular to the limb.
\paragraph{Spectra parallel to the limb -- horizontal variation}
Figure \ref{fig6} shows the observed spectra on cuts parallel to the limb (some examples of the corresponding heights above the limb are marked in the {\em rightmost panel} of Fig.~\ref{fig3}). 
The maximal intensity at each height is denoted 
in the {\em lower right corner}. One outstanding pattern throughout all heights are the Doppler excursions of the absorption core. For many spatial locations, the blue and  red shifts can be traced through all heights. They are especially visible in the height range from 2.5 to 4 Mm, yielding a clear zig-zag pattern. At lower layers, they can be seen in the intensity difference between the H$_{\rm 2V}$ and H$_{\rm 2R}$ emission peaks which brighten and darken 
with the shifts of the absorption core ($d$=1 to 2 Mm, $x\sim$20$^{\prime\prime}$). This phenomenon has often been called line tilt \citep[e.g.,][]{beckers68,pasachoff+etal1968}. An extensive discussion of possible explanations of these line tilts is given by \citet{rompolt1975}. 

Six examples of prominent Doppler excursions are marked in the spectra of the
cut parallel to the limb at 2.5 Mm height ({\em second column, bottom row}). The
resulting line tilt corresponds to a shift of the central absorption core that
partially suppresses one of the emission peaks while enhancing the other
one. The variation of the strength of the emission peaks is even more
pronounced at about $x=20^{\prime\prime}$. The evolution of the spectra from
a red shift to a blue shift and back takes part on a typical scale of 4\,--\,6
profiles ($\equiv$1.2 to 1.8$^{\prime\prime}$); Figure \ref{fig7_app} shows
the individual spectra around the positions marked in Fig.~\ref{fig6}. Figure
\ref{fig11} displays the appearance of the line-core shifts in the velocity
proxies and intensity bands. They are clearly mirrored in the intensity
difference of the H$_{\rm 2V}$ and H$_{\rm 2R}$ emission peaks and the
line-core velocity ({\em top panel}). The velocity amplitude is about $\pm$4
kms$^{-1}$. There is a trend for local intensity maxima of the H-index, which
is insensitive to the Doppler shifts, to occur between the maximal and minimal
velocities ({\em dashed lines} in the {\em bottom panel}). 
\begin{figure}
\centerline{\resizebox{8.cm}{!}{\includegraphics{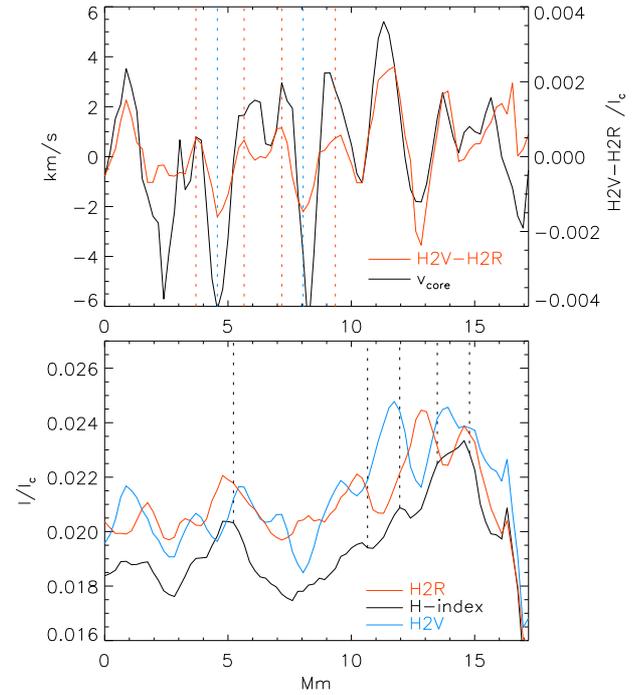}}}
\caption{Appearance of the Doppler excursions at 2.5 Mm in intensity and velocity. {{\em Top}: 
velocity proxies. {\em Red}: difference of H$_{\rm 2V}$ and H$_{\rm 2R}$. {\em Black}: position of the line core. 
The {\em blue} and {\em red dashed vertical lines} denote the locations of blue/red shifts. {\em Bottom}: 
intensity of H$_{\rm 2V}$ ({\em blue}),  H$_{\rm 2R}$ ({\em red}), and the H-index ({\em black}). 
The {\em vertical black dashed lines} indicate local maxima of the H-index. The abscissa is in Mm.} \label{fig11}}
\end{figure}
\begin{figure}[hb]
\centerline{\resizebox{8.8cm}{!}{\includegraphics{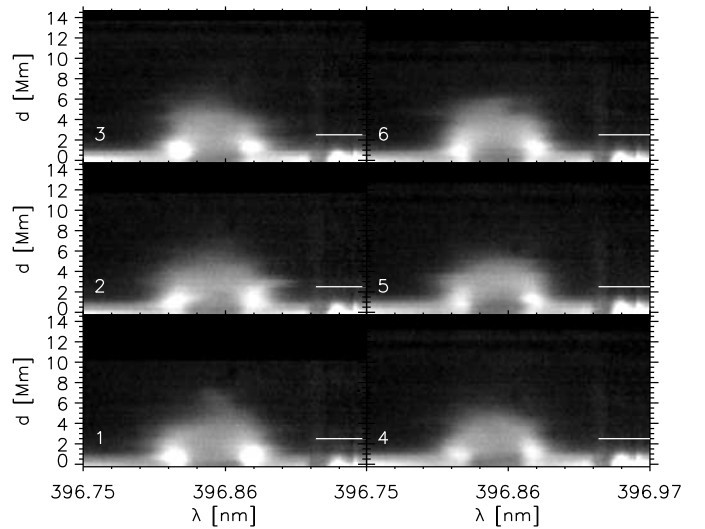}}}
\caption{Spectra corresponding to the cuts 1 to 6 marked in Fig.~\ref{fig2b}. The limb is at bottom, the $y$-axis 
gives the distance $d$ to the limb. The {\em horizontal white bars} at the right indicate a height of 2.5 Mm. 
The display is thresholded at 3\% of $I_c$ for better visibility.\label{fig13}}
\end{figure} 
\paragraph{Spectra perpendicular to the limb -- vertical variation} 
The spectra shown in the following Figs.~\ref{fig13} and \ref{fig17} were taken along the cuts marked in 
the {\em top left} panels of Fig.~\ref{fig2b}. The six rays perpendicular to the limb, labeled 1 to 6, were chosen 
to intersect the line of 2.5 Mm distance to the limb such that they correspond to the locations of the large Doppler 
excursions marked in Fig.~\ref{fig6}. The spectra along these cuts are shown in Fig.~\ref{fig13}. 
The Doppler excursions are seen on the one hand as bright ``branches'' to the blue and red; on the other hand, 
they also appear in the variation of the relative intensities of H$_{\rm 2V}$ and  H$_{\rm 2R}$. In some cases, 
the Doppler shifts show an oscillatory pattern not only in the horizontal direction (as seen by the variation 
between ray 1, 2, and 3), but in the vertical direction as well, e.g., for the case of ray 2, giving the spectra 
a faint similarity with a Christmas tree. The individual spectra of cut no.~2 are shown in Fig.~\ref{fig22}.

The other cuts were taken along the slit for two reasons. First, the
spectra of a slit position are taken simultaneously, so only a temporal
evolution faster than the 13-sec integration time can play a role. Second,
many of the elongated structures seen in for instance the IW1 map of
Fig.~\ref{fig2b} are rather oriented along the slit than perpendicular to the
limb. For the cuts along the slit, we always marked one of the scan steps with its number in Fig.~\ref{fig2b}, to be able to relate the shape of the spectra on that location to a specific prominent feature in the 2-D maps of the FOV. Because the general patterns in these spectra are similar to those of Fig.~\ref{fig13}, we moved them to the appendix. Repeated Doppler excursions to the red and the blue along the vertical direction can be seen in several examples; the largest Doppler shifts are usually seen at some distance (about 2-6 Mm) above the limb. The Doppler excursions are coherent over some scan positions, e.g., for the scan steps 65 to 68 at $y=13^{\prime\prime}$ (Fig.~\ref{fig17}), which also implies a lower limit for the duration of about 50 sec in addition to the spatial extent of 1-2$^{\prime\prime}$. Some features like the strong brightening of H$_{\rm 2R}$ in scan step 69 at $y=11^{\prime\prime}$ (also in Fig.~\ref{fig17}) fade away (toward subsequent steps) or significantly change shape (toward previous scan steps) over 2-3 steps ($\equiv$ 0\farcs6-0\farcs9) only, with 0\farcs6 being the technical lowermost possible limit for the resolution. The individual spectra of the scan step 66 are shown in Fig.~\ref{fig23}.
\begin{figure}
\centerline{\resizebox{8.8cm}{!}{\includegraphics{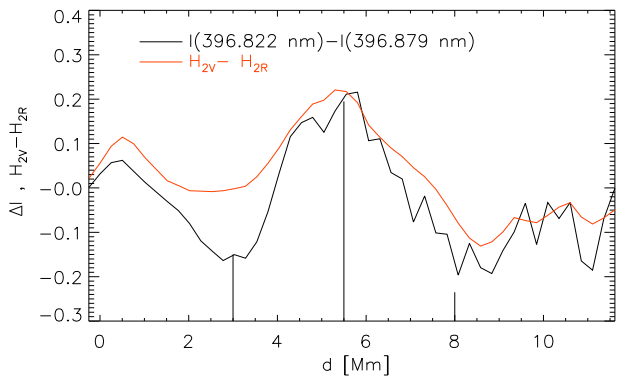}}}
\caption{Vertical variation of intensities along cut no.~2. {\em Black}: difference of intensities at 396.822\thinspace nm and 396.879\thinspace nm. {\em Red}: difference of H$_{\rm 2V}$ and H$_{\rm 2R}$. \label{fig14}}
\end{figure}

The spatial extent of the Doppler variations along the local vertical is shown
in Fig.~\ref{fig14} using the profiles of cut no.~2 (cf.~also the map of H$_{\rm 2V}$/H$_{\rm 2R}$ in Fig.~\ref{fig3} at $(x,y)=(6^{\prime\prime},50^{\prime\prime})$\,). We calculated the
intensity difference of H$_{\rm 2V}$ and H$_{\rm 2R}$ in each profile ({\em
  red line}), and also created a single-wavelength ``Dopplergram'' by
subtracting $I(\lambda = 396.879$\thinspace nm) from $I(\lambda =
396.822$\thinspace nm) for comparison ({\em black line}). The latter two
wavelengths sample only the faint ``branches'' to the blue and red and
  are therefore partly insensitive to the location of the central absorption core measuring shifts of the complete emission pattern, but basically yield the same spatial variation as the emission peaks. Owing to the low intensity of the ``branches'', the other velocity proxies do not capture the Doppler shifts well for these profiles. To improve the visibility of the Doppler shifts, we removed the intensity decrease with distance from the limb by dividing the differences with the sum of the two terms. The recursive Doppler oscillations then occur with a separation of about 2.5-3 Mm ({\em black vertical lines}) along the local vertical. It cannot be excluded, however, that the LOS crossed more than one distinct structure and sampled its contribution to the spectra. The repetitive Doppler excursions perpendicular to the limb are in general less pronounced than the variation on cuts parallel to the limb.
\subsection{Statistics of Doppler shifts}
Because the Ca line profile looses its central absorption core at a height of 5 Mm above the limb and other LOS velocity proxies like the difference of  H$_{\rm 2V}$ and H$_{\rm 2R}$ do not yield a quantitative velocity measure, we restricted the analysis of the statistics of Doppler shifts to a maximum height of 5 Mm. The unsigned velocity ranges up to about 15 kms$^{-1}$, with a mean value of 3.2 kms$^{-1}$. The root-mean-square (rms) for the signed velocity is about 4 kms$^{-1}$, which agrees with the 3.8 kms$^{-1}$ given by BE68 for a height of 6 Mm. We find that the rms velocity of the photospheric blends and the Ca core increases toward the limb, similar to the behavior of the line width shown in Fig.~\ref{fig42}.
\section{Summary \label{sect_disc}}
We investigated a set of spectra of \ion{Ca}{ii}\,H inside a FOV that extended
some 20$^{\prime\prime}$ beyond the solar limb. We estimate that we
achieved a spatial resolution of about 1\,--\,1.5$^{\prime\prime}$ with permanent image
stabilization to better than about 0\farcs5. The data set allowed us to retrieve average off-limb Ca spectra with a well-defined limb distance. The average spectra show a smoothly increasing width of the Ca line core up to about 2-3 Mm height above the limb. At a height of 5 Mm, the line width starts to reduce strongly. The emission peaks are generally symmetric. Above a limb distance of 5\,Mm, the profile shape is similar to a single Gaussian. Wavelengths in and near the core show a local intensity maximum above the limb, whose height decreases with separation from the very core, as expected from intensity contribution functions. The found variation with limb distance of the intensity, the integrated emission, and the line width should be reproduced by theoretical models, providing boundary conditions for, e.g., the microturbulent velocity and the temperature stratification of the chromosphere. 

Individual profiles from beyond the limb show ubiquitously large Doppler shifts, visible either as a clear shift of the absorption core or as a pronounced asymmetry of H$_{\rm 2V}$ and H$_{\rm 2R}$. On cuts parallel to the limb, the sign of the velocity reverses on the smallest spatial scales resolved in the data ($\sim 1.5^{\prime\prime}$). The velocity amplitude as determined from the position of the absorption core is about $\pm4$\thinspace kms$^{-1}$. Cuts perpendicular to the limb or along the slit show in some cases a similar pattern of sign reversal of the velocity. The changes of the velocity sign often extend over a few arcseconds in the direction perpendicular to the limb and flank slight enhancements of the total emission as given by the wavelength-integrated H-index. 
\section{Discussion \label{sect_concl}}
We have obtained medium-resolution slit spectra of \ion{Ca}{ii} H near and above the limb. The series of disk center observations with identical exposure times obtained during the observing day allowed us to obtain an absolute intensity calibration whose accuracy is only limited by the  applied stray-light corrections. The coefficient $\alpha$ of the stray-light correction directly scales the total intensity by the subtraction of the average profile. This affects the spectra on the disk most strongly because beyond the limb the stray-light coefficient is rapidly decreasing (Fig.~\ref{fig_1a}). Beyond the limb, the intrinsic emission in the line core with an intensity of about 1\,\% at a height of 5\,Mm (cf.~Fig.~\ref{fig5}) dominates over at least the residual stray light after the correction. At heights above about 9\,Mm, the intensity in the line wing and the line core equalize. The remaining intensity in the core of the line (and also all other wavelengths) is presumably only caused by the residual stray-light level because the chromosphere itself has a finite extent of a few Mm. We tested different values of $\alpha$ and of the off-limb correction, but the resulting intensity in the off-disk emission was roughly independent of the exact way of stray-light correction \citep[see][submitted to A\&A]{beck+rezaei2010a}. 

Our spectroscopic observations were obtained with real-time image stabilization and correction by an AO system, with a residual image motion below 1$^{\prime\prime}$ (see Appendix \ref{spat_res_sect}). For this, no dedicated device like a limb-tracker as suggested by \citet{pasachoff+etal09} is necessary, it is sufficient if adaptive optics and the spatial scanning for a spectrograph or the selection of the FOV to be observed, respectively, are independent of each other\footnote{This is a crucial point for any AO-supported observations near the limb that deserves attention in telescope designs.}. The image stability and the complete coverage of the height range allowed us to sample the spectra continuously with height with a sampling of about 0\farcs3. In some older observations \citep[e.g.,][PA68]{zirker1962,pasachoff+etal1968}, the spatial sampling was significantly lower or even reduced to two height levels only in the latter case. The Doppler shifts have a clear signature in our off-limb spectra, appearing in two different shapes: displacements of the absorption core relative to the emission peaks, and shifts of the complete profile seen as faint ``whiskers'' to the red and blue of the line core. In the spectra used by PA68, no displacements of the absorption core were seen (compare their Fig.~11 with Fig.~\ref{fig7_app}). We relate this to the spatial resolution achieved in our observations because in the average spectra or for a spatial smearing over 2-3$^{\prime\prime}$ the oppositely directed Doppler shifts cancel out again. 

The observed width of the \ion{Ca}{ii} H (or K) line can presumably not be caused only by different Doppler shifts of unresolved sub-structures \citep{athay1961,athay+bessey1964}. Many multi-line observations of limb structures \citep[e.g.,][]{krat+krat1971,alissandrakis1973,athay+bessey1964,pasachoff+etal1968,socasnavarro+elmore2005} also implied that other processes than thermal Doppler broadening have to be present because the observed line widths of Ca and other elements could not be reproduced assuming the same temperature and its corresponding thermal broadening. In our case, the line width of the average off-limb profiles shows a smooth variation with an abrupt change at about 5\,Mm limb distance. The height of this transition is fairly constant all across the FOV in the spatially resolved spectra. This suggests an optical depth effect caused by absorbing matter in front of an emission background. Eventually the difference between the two type of ``wide'' (large line width) and ``narrow'' (small line width) spicules defined by BE68 marks the difference between optically thin and thick structures, where only for the optically thick structures a sufficient self-absorption is present.  

The origin of the observed relation (Fig.~\ref{fig11}) between
enhancements of the wavelength-integrated H-index -- which to first order is
insensitive to Doppler shifts of the line core -- and the oppositely directed
LOS velocities that flank them is unclear at first. The pattern is not
  only seen at the height of 2.5 Mm shown in Fig.~\ref{fig11}, but continues
  over a height range of about 5 Mm (cf.~the maps of the H-index in Fig.~\ref{fig2b} and that of H$_{\rm 2R}$/H$_{\rm 2V}$ in Fig.~\ref{fig3} at $(x,y)=(20^{\prime\prime}, 62^{\prime\prime})$\,).
\begin{figure}
$ $\\
$ $\\
\centerline{\resizebox{8.8cm}{!}{\includegraphics{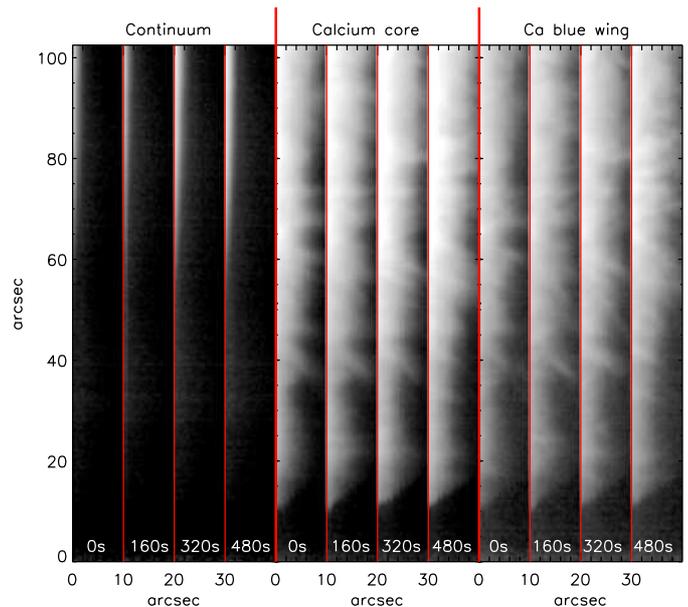}}}
\caption{Example of a time-series of \ion{Ca}{ii} H spectra taken in July 2010. {\em Left}: continuum intensity. {\em Middle}: line-core intensity. {\em Right}: intensity to the blue of the line core. Exposure time per scan step was 6 seconds. All images are displayed in logarithmic scale.\label{fig_dangerous}}
\end{figure}

There are several possible explanations for the relation between the
  velocity and intensity pattern. One is that the corresponding solar
  structures are rotating as a rigid body, which was invoked recently for
  explaining the observation of a macro-spicule by
  \citet{kamio+etal2010}. Other explanations are waves propagating along the
  field lines of an intrinsically twisted flux tube \citep{avery1970} or waves
  traveling at an angle to the orientation of the tube axis, the twisting motion of field lines as suggested by 
the high-resolution, high-cadence Ca imaging data from HINODE \citep{suematsu+etal2008}, or helical kink 
waves \citep{zaqarashvili+etal2008}. The Doppler shift pattern caused by a helical wave propagation along 
a central flux tube would presumably be different from the sausage soliton solution proposed by \citet{zaqarashvili+etal2010}, which would have a maximum velocity amplitude at the center of the brightening instead of the velocity sign reversal on the sides. In view of the limited spatial resolution and the low statistics of our data, we refrain from trying to give a definite answer, however. 

Our observations have one drawback for the interpretation of the findings. The orientation of the slit was to first order perpendicular to the limb. Even with the image correction by AO, this increases the stray-light contamination in the off-limb spectra because the on-disk spectra have an intensity higher by an order of magnitude. Observations with the slit parallel to the limb are therefore to be preferred, possibly also a set of repeated scans to allow for a more reliable identification of features. Such data were obtained during an observation 
campaign in July 2010, but their analysis is still pending. Figure \ref{fig_dangerous} shows an example of the improvement in spatial resolution that is to be expected; the corresponding data have {\em not} been corrected for any stray light yet. 

The \ion{Ca}{ii}\,H line is less accessible from the theoretical side than the \ion{Ca}{ii}\,IR lines because of the partial redistribution effects inside the spectral line \citep{heggland+etal2009,leenaarts+etal2009}, but it still is easier to handle than H$_\alpha$ \citep{carlsson+stein2002,leenaarts+etal2007}. Because many recent results are based on data obtained with the broad-band Ca filter imaging of HINODE without spectral resolution, it seems worthwhile to cross-check the features in these observations with spectra, eventually from a new solar space mission like Solar-C. A spectral synthesis from simulations like described in \citet{martinez+etal2009} would also be a possibility; or a direct comparison with the synthetic spectra recently presented in \citet{judge+carlsson2010}.

Our data set (41 MB) is available online\footnote{http://www3.kis.uni-freiburg.de/$\sim$rrezaei/spiculedata/} for free use. 
This will hopefully  raise some more interest in the development of analysis methods for the \ion{Ca}{ii}\,H line or obtaining 
high-S/N slit spectra even in modern times, despite the claim in the introduction of \citet{leenaarts+etal2010} that 
only 2D imaging spectroscopy and spectropolarimetry is ``{\em adequate}'' for studying the chromosphere, 
or the similar one in the conclusions of \citet{judge+etal2010}.
\section{Conclusions\label{sect_concll}}
Observations near the solar limb with a real-time correction by adaptive optics are possible with medium to high spatial resolution when the AO system allows one to relocated the observed FOV from the AO lock point. Slit-spectrograph observations with their full line-spectra provide the information on both the thermal and dynamical state of the atmosphere, where the information on the Doppler shifts allows different scenarios such as helical or sausage wave modes to be distinguished. The average off-limb spectra can finally be used to test static atmosphere models because above the limb it is ensured that they only sample chromospheric height layers.
\begin{acknowledgements}
The VTT is operated by the Kiepenheuer-Institut f\"ur Sonnenphysik (KIS) at the Spanish Observatorio del Teide of 
the Instituto de Astrof\'{\i}sica de Canarias (IAC). The POLIS instrument has been a joint development of 
the High Altitude Observatory (Boulder, USA) and the KIS. C.B.~acknowledges partial support by the Spanish Ministry 
of Science and Innovation through project AYA 2007-63881. R.R.~acknowledges partial support by the Deutsche 
Forschungsgemeinschaft under grant SCHM 1168/8-2. 
We thank B.~Lites for discussions on the topic of stray light.
\end{acknowledgements}
\bibliographystyle{aa}
\bibliography{references_luis_mod2}

\begin{thebibliography}{78}
\expandafter\ifx\csname natexlab\endcsname\relax\def\natexlab#1{#1}\fi

\bibitem[{{Alissandrakis}(1973)}]{alissandrakis1973}
{Alissandrakis}, C.~E. 1973, \solphys, 32, 345

\bibitem[{{Athay}(1961)}]{athay1961}
{Athay}, R.~G. 1961, \apj, 134, 756

\bibitem[{{Athay} \& {Bessey}(1964)}]{athay+bessey1964}
{Athay}, R.~G. \& {Bessey}, R.~J. 1964, \apj, 140, 1174

\bibitem[{{Avery}(1970)}]{avery1970}
{Avery}, L.~W. 1970, \solphys, 13, 301

\bibitem[{{Balasubramaniam} {et~al.}(2004){Balasubramaniam}, {Christopoulou},
  \& {Uitenbroek}}]{bala+etal2004}
{Balasubramaniam}, K.~S., {Christopoulou}, E.~B., \& {Uitenbroek}, H. 2004,
  \apj, 606, 1233

\bibitem[{{Beck} {et~al.}(2009){Beck}, {Khomenko}, {Rezaei}, \&
  {Collados}}]{beck+etal2009}
{Beck}, C., {Khomenko}, E., {Rezaei}, R., \& {Collados}, M. 2009, \aap, 507,
  453

\bibitem[{{Beck} {et~al.}(2011){Beck}, {Rezaei}, \&
  {Fabbian}}]{beck+rezaei2010a}
{Beck}, C., {Rezaei}, R., \& {Fabbian}, D. 2011, {\em Stray-light contamination
  and spatial de-convolution of slit-spectrograph observations}, submitted to
  A\&A

\bibitem[{{Beck} {et~al.}(2005{\natexlab{a}}){Beck}, {Schlichenmaier},
  {Collados}, {Bellot Rubio}, \& {Kentischer}}]{beck+etal2005a}
{Beck}, C., {Schlichenmaier}, R., {Collados}, M., {Bellot Rubio}, L., \&
  {Kentischer}, T. 2005{\natexlab{a}}, \aap, 443, 1047

\bibitem[{{Beck} {et~al.}(2005{\natexlab{b}}){Beck}, {Schmidt}, {Kentischer},
  \& {Elmore}}]{beck+etal2005b}
{Beck}, C., {Schmidt}, W., {Kentischer}, T., \& {Elmore}, D.
  2005{\natexlab{b}}, \aap, 437, 1159

\bibitem[{{Beck} {et~al.}(2008){Beck}, {Schmidt}, {Rezaei}, \&
  {Rammacher}}]{beck+etal2008}
{Beck}, C., {Schmidt}, W., {Rezaei}, R., \& {Rammacher}, W. 2008, \aap, 479,
  213

\bibitem[{{Beckers}(1968)}]{beckers68}
{Beckers}, J.~M. 1968, \solphys, 3, 367 (BE68)

\bibitem[{{Biermann}(1948)}]{biermann_48}
{Biermann}, L. 1948, Zeitschrift fur Astrophysik, 25, 161

\bibitem[{{Carlsson} \& {Stein}(1997)}]{carlsson+stein1997}
{Carlsson}, M. \& {Stein}, R.~F. 1997, \apj, 481, 500

\bibitem[{{Carlsson} \& {Stein}(2002)}]{carlsson+stein2002}
{Carlsson}, M. \& {Stein}, R.~F. 2002, \apj, 572, 626

\bibitem[{{Cauzzi} {et~al.}(2008){Cauzzi}, {Reardon}, {Uitenbroek},
  {Cavallini}, {Falchi}, {Falciani}, {Janssen}, {Rimmele}, {Vecchio}, \&
  {W{\"o}ger}}]{cauzzi+etal2008}
{Cauzzi}, G., {Reardon}, K.~P., {Uitenbroek}, H., {et~al.} 2008, \aap, 480, 515

\bibitem[{{Centeno} {et~al.}(2010){Centeno}, {Trujillo Bueno}, \& {Asensio
  Ramos}}]{centeno+etal2010}
{Centeno}, R., {Trujillo Bueno}, J., \& {Asensio Ramos}, A. 2010, \apj, 708,
  1579

\bibitem[{{Cram} \& {Dame}(1983)}]{cram+dame1983}
{Cram}, L.~E. \& {Dame}, L. 1983, \apj, 272, 355

\bibitem[{{Cram} {et~al.}(1980){Cram}, {Lites}, \& {Rutten}}]{cram+etal1980}
{Cram}, L.~E., {Lites}, B.~W., \& {Rutten}, R.~J. 1980, \apj, 241, 374

\bibitem[{{Dunn} {et~al.}(1968){Dunn}, {Evans}, {Jefferies}, {Orrall}, {White},
  \& {Zirker}}]{dunn+etal1968}
{Dunn}, R.~B., {Evans}, J.~W., {Jefferies}, J.~T., {et~al.} 1968, \apjs, 15,
  275

\bibitem[{{Engvold} \& {Halvorsen}(1973)}]{engvold+halvorsen1973}
{Engvold}, O. \& {Halvorsen}, H.~D. 1973, \solphys, 28, 23

\bibitem[{{Fontenla} {et~al.}(2006){Fontenla}, {Avrett}, {Thuillier}, \&
  {Harder}}]{fontenla_etal_06}
{Fontenla}, J.~M., {Avrett}, E., {Thuillier}, G., \& {Harder}, J. 2006, \apj,
  639, 441

\bibitem[{{Fontenla} {et~al.}(2007){Fontenla}, {Balasubramaniam}, \&
  {Harder}}]{fontenla_etal_07}
{Fontenla}, J.~M., {Balasubramaniam}, K.~S., \& {Harder}, J. 2007, in The
  Physics of Chromospheric Plasma, ed. P.~{Heinzel}, I.~{Dorotovic}, \& R.~R.
  J., 499

\bibitem[{{Fossum} \& {Carlsson}(2005)}]{fossum+carlsson2005}
{Fossum}, A. \& {Carlsson}, M. 2005, \nat, 435, 919

\bibitem[{{Haberreiter} {et~al.}(2010){Haberreiter}, {Finsterle}, {McIntosh},
  \& {Wedemeyer-Boehm}}]{haberreiter+etal2010}
{Haberreiter}, M., {Finsterle}, W., {McIntosh}, S., \& {Wedemeyer-Boehm}, S.
  2010, \memsai, 81, 782

\bibitem[{{Heggland} {et~al.}(2009){Heggland}, {De Pontieu}, \&
  {Hansteen}}]{heggland+etal2009}
{Heggland}, L., {De Pontieu}, B., \& {Hansteen}, V.~H. 2009, \apj, 702, 1

\bibitem[{{Jebsen} \& {Mitchell}(1978)}]{jebsen+mitchell1978}
{Jebsen}, D.~E. \& {Mitchell}, Jr., W.~E. 1978, \solphys, 57, 309

\bibitem[{{Judge} \& {Carlsson}(2010)}]{judge+carlsson2010}
{Judge}, P.~G. \& {Carlsson}, M. 2010, \apj, 719, 469

\bibitem[{{Judge} {et~al.}(2010){Judge}, {Tritschler}, {Uitenbroek}, {Reardon},
  {Cauzzi}, \& {de Wijn}}]{judge+etal2010}
{Judge}, P.~G., {Tritschler}, A., {Uitenbroek}, H., {et~al.} 2010, \apj, 710,
  1486

\bibitem[{{Kalkofen}(1996)}]{kalkofen1996}
{Kalkofen}, W. 1996, \apjl, 468, L69

\bibitem[{{Kamio} {et~al.}(2010){Kamio}, {Curdt}, {Teriaca}, {Inhester}, \&
  {Solanki}}]{kamio+etal2010}
{Kamio}, S., {Curdt}, W., {Teriaca}, L., {Inhester}, B., \& {Solanki}, S.~K.
  2010, \aap, 510, L1

\bibitem[{{Kosugi} {et~al.}(2007){Kosugi}, {Matsuzaki}, {Sakao}, {Shimizu},
  {Sone}, {Tachikawa}, {Hashimoto}, {Minesugi}, {Ohnishi}, {Yamada}, {Tsuneta},
  {Hara}, {Ichimoto}, {Suematsu}, {Shimojo}, {Watanabe}, {Shimada}, {Davis},
  {Hill}, {Owens}, {Title}, {Culhane}, {Harra}, {Doschek}, \&
  {Golub}}]{kosugi+etal2007}
{Kosugi}, T., {Matsuzaki}, K., {Sakao}, T., {et~al.} 2007, \solphys, 243, 3

\bibitem[{{Krat} \& {Krat}(1971)}]{krat+krat1971}
{Krat}, V.~A. \& {Krat}, T.~V. 1971, \solphys, 17, 355

\bibitem[{{Kurucz} {et~al.}(1984){Kurucz}, {Furenlid}, {Brault}, \&
  {Testerman}}]{kurucz+etal1984}
{Kurucz}, R.~L., {Furenlid}, I., {Brault}, J., \& {Testerman}, L. 1984, {Solar
  flux atlas from 296 to 1300 nm} (NSO, Sunspot, NM)

\bibitem[{{Leenaarts} {et~al.}(2009){Leenaarts}, {Carlsson}, {Hansteen}, \&
  {Rouppe van der Voort}}]{leenaarts+etal2009}
{Leenaarts}, J., {Carlsson}, M., {Hansteen}, V., \& {Rouppe van der Voort}, L.
  2009, \apjl, 694, L128

\bibitem[{{Leenaarts} {et~al.}(2007){Leenaarts}, {Carlsson}, {Hansteen}, \&
  {Rutten}}]{leenaarts+etal2007}
{Leenaarts}, J., {Carlsson}, M., {Hansteen}, V., \& {Rutten}, R.~J. 2007, \aap,
  473, 625

\bibitem[{{Leenaarts} {et~al.}(2010){Leenaarts}, {Rutten}, {Reardon},
  {Carlsson}, \& {Hansteen}}]{leenaarts+etal2010}
{Leenaarts}, J., {Rutten}, R.~J., {Reardon}, K., {Carlsson}, M., \& {Hansteen},
  V. 2010, \apj, 709, 1362

\bibitem[{{Linsky} \& {Avrett}(1970)}]{linsky+avrett1970}
{Linsky}, J.~L. \& {Avrett}, E.~H. 1970, \pasp, 82, 169

\bibitem[{{Lites}(1974)}]{lites1974}
{Lites}, B.~W. 1974, \aap, 33, 363

\bibitem[{{Lites} {et~al.}(1993){Lites}, {Rutten}, \&
  {Kalkofen}}]{lites+etal1993}
{Lites}, B.~W., {Rutten}, R.~J., \& {Kalkofen}, W. 1993, \apj, 414, 345

\bibitem[{{Makita}(2003)}]{makita2003}
{Makita}, M. 2003, Publications of the National Astronomical Observatory of
  Japan, 7, 1

\bibitem[{{Martinez Pillet} {et~al.}(1990){Martinez Pillet}, {Ruiz Cobo}, \&
  {Vazquez}}]{martinezpillet+etal1990}
{Martinez Pillet}, V., {Ruiz Cobo}, B., \& {Vazquez}, M. 1990, \solphys, 125,
  211

\bibitem[{{Mart{\'{\i}}nez-Sykora} {et~al.}(2009){Mart{\'{\i}}nez-Sykora},
  {Hansteen}, {DePontieu}, \& {Carlsson}}]{martinez+etal2009}
{Mart{\'{\i}}nez-Sykora}, J., {Hansteen}, V., {DePontieu}, B., \& {Carlsson},
  M. 2009, \apj, 701, 1569

\bibitem[{{Mattig}(1971)}]{mattig1971}
{Mattig}, W. 1971, \solphys, 18, 434

\bibitem[{{Mattig}(1983)}]{mattig1983}
{Mattig}, W. 1983, \solphys, 87, 187

\bibitem[{{Narain} \& {Ulmschneider}(1996)}]{nar_ulm_96}
{Narain}, U. \& {Ulmschneider}, P. 1996, Space Science Reviews, 75, 453

\bibitem[{{Oranje}(1983)}]{oranje1983}
{Oranje}, B.~J. 1983, \aap, 124, 43

\bibitem[{{Pasachoff} {et~al.}(2009){Pasachoff}, {Jacobson}, \&
  {Sterling}}]{pasachoff+etal09}
{Pasachoff}, J.~M., {Jacobson}, W.~A., \& {Sterling}, A.~C. 2009, \solphys,
  260, 59

\bibitem[{{Pasachoff} {et~al.}(1968){Pasachoff}, {Noyes}, \&
  {Beckers}}]{pasachoff+etal1968}
{Pasachoff}, J.~M., {Noyes}, R.~W., \& {Beckers}, J.~M. 1968, \solphys, 5, 131

\bibitem[{{Penn} \& {Jones}(1996)}]{penn+jones1996}
{Penn}, M.~J. \& {Jones}, H.~P. 1996, \solphys, 168, 19

\bibitem[{{Pietarila} {et~al.}(2007){Pietarila}, {Socas-Navarro}, \&
  {Bogdan}}]{pietarila+etal2007}
{Pietarila}, A., {Socas-Navarro}, H., \& {Bogdan}, T. 2007, \apj, 663, 1386

\bibitem[{{Rezaei} {et~al.}(2008){Rezaei}, {Bruls}, {Schmidt}, {Beck},
  {Kalkofen}, \& {Schlichenmaier}}]{rezaei+etal2008}
{Rezaei}, R., {Bruls}, J.~H.~M.~J., {Schmidt}, W., {et~al.} 2008, \aap, 484,
  503

\bibitem[{{Rezaei} {et~al.}(2007){Rezaei}, {Schlichenmaier}, {Beck}, {Bruls},
  \& {Schmidt}}]{rezaei+etal2007}
{Rezaei}, R., {Schlichenmaier}, R., {Beck}, C.~A.~R., {Bruls}, J.~H.~M.~J., \&
  {Schmidt}, W. 2007, \aap, 466, 1131

\bibitem[{{Rompolt}(1975)}]{rompolt1975}
{Rompolt}, B. 1975, \solphys, 41, 329

\bibitem[{{Rutten}(2007)}]{rutten2007}
{Rutten}, R.~J. 2007, in The Physics of Chromospheric Plasmas, ed. {P.~Heinzel,
  I.~Dorotovi{\v c}, \& R.~J.~Rutten}, ASP Conf.~Series, 368, 27

\bibitem[{{Rutten}(2010)}]{rutten2010}
{Rutten}, R.~J. 2010, in Recent Advances in Spectroscopy: Theoretical,
  Astrophysical and Experimental Perspectives, ed. {R.~K.~Chaudhuri,
  M.~V.~Mekkaden, A.~V.~Raveendran, \& A.~Satya Narayanan} (Springer, Berlin),
  163--175

\bibitem[{{Rutten} \& {Stencel}(1980)}]{rutten+stencel1980}
{Rutten}, R.~J. \& {Stencel}, R.~E. 1980, \aaps, 39, 415

\bibitem[{{S{\'a}nchez-Andrade Nu{\~n}o} {et~al.}(2007){S{\'a}nchez-Andrade
  Nu{\~n}o}, {Centeno}, {Puschmann}, {Trujillo Bueno}, {Blanco
  Rodr{\'{\i}}guez}, \& {Kneer}}]{sanchezandrade+etal2007}
{S{\'a}nchez-Andrade Nu{\~n}o}, B., {Centeno}, R., {Puschmann}, K.~G., {et~al.}
  2007, \aap, 472, L51

\bibitem[{{Schmidt} {et~al.}(1994){Schmidt}, {Knoelker}, \& {Westendrop
  Plaza}}]{schmidt+etal1994}
{Schmidt}, W., {Knoelker}, M., \& {Westendrop Plaza}, C. 1994, \aap, 287, 229

\bibitem[{{Schr\"oter} {et~al.}(1985){Schr\"oter}, {Soltau}, \& {Wiehr}}]{vtt}
{Schr\"oter}, E.~H., {Soltau}, D., \& {Wiehr}, E. 1985, Vistas in Astronomy,
  28, 519

\bibitem[{{Shoji} {et~al.}(2010){Shoji}, {Nishikawa}, {Kitai}, \&
  {Ueno}}]{shoji+etal2010}
{Shoji}, M., {Nishikawa}, T., {Kitai}, R., \& {Ueno}, S. 2010, \pasj, 62, 927

\bibitem[{{Sivaraman} {et~al.}(1996){Sivaraman}, {Gupta}, \&
  {Kariyappa}}]{sivaraman+etal1996}
{Sivaraman}, K.~R., {Gupta}, S.~S., \& {Kariyappa}, R. 1996, \solphys, 163, 93

\bibitem[{{Sivaraman} {et~al.}(1987){Sivaraman}, {Singh}, {Bagare}, \&
  {Gupta}}]{sivaraman+etal1987}
{Sivaraman}, K.~R., {Singh}, J., {Bagare}, S.~P., \& {Gupta}, S.~S. 1987, \apj,
  313, 456

\bibitem[{{Socas-Navarro}(2005)}]{navarro2005}
{Socas-Navarro}, H. 2005, \apjl, 633, L57

\bibitem[{{Socas-Navarro} \& {Elmore}(2005)}]{socasnavarro+elmore2005}
{Socas-Navarro}, H. \& {Elmore}, D. 2005, \apjl, 619, L195

\bibitem[{{Stencel}(2009)}]{stencel2009}
{Stencel}, R.~E. 2009, in The Biggest, Baddest, Coolest Stars, ed.
  {D.~G.~Luttermoser, B.~J.~Smith, \& R.~E.~Stencel}, ASP Conf.~Series, 412,
  251

\bibitem[{{Suematsu} {et~al.}(2008){Suematsu}, {Ichimoto}, {Katsukawa},
  {Shimizu}, {Okamoto}, {Tsuneta}, {Tarbell}, \& {Shine}}]{suematsu+etal2008}
{Suematsu}, Y., {Ichimoto}, K., {Katsukawa}, Y., {et~al.} 2008, in First
  Results From Hinode, ed. {S.~A.~Matthews, J.~M.~Davis, \& L.~K.~Harra}, ASP
  Conf.~Series, 397, 27

\bibitem[{{Teplitskaja} \& {Efendieva}(1973)}]{teplitskaja+efendieva1973}
{Teplitskaja}, R.~B. \& {Efendieva}, S.~A. 1973, \solphys, 28, 369

\bibitem[{{Vernazza} {et~al.}(1981){Vernazza}, {Avrett}, \&
  {Loeser}}]{vernazza+etal1981}
{Vernazza}, J.~E., {Avrett}, E.~H., \& {Loeser}, R. 1981, \apjs, 45, 635

\bibitem[{{von der L\"uhe} {et~al.}(2003){von der L\"uhe}, {Soltau},
  {Berkefeld}, \& {Schelenz}}]{vdluehe+etal2003}
{von der L\"uhe}, O., {Soltau}, D., {Berkefeld}, T., \& {Schelenz}, T. 2003, in
  Innovative Telescopes and Instrumentation for Solar Astrophysics, ed. S.~Keil
  \& S.~Avakyan, Proceedings of the SPIE, 4853, 187

\bibitem[{{Watanabe} \& {Steenbock}(1986)}]{watanabe+steenbock1986}
{Watanabe}, T. \& {Steenbock}, W. 1986, \aap, 165, 163

\bibitem[{{Wedemeyer} {et~al.}(2004){Wedemeyer}, {Freytag}, {Steffen},
  {Ludwig}, \& {Holweger}}]{wedemeyer+etal2004}
{Wedemeyer}, S., {Freytag}, B., {Steffen}, M., {Ludwig}, H.-G., \& {Holweger},
  H. 2004, \aap, 414, 1121

\bibitem[{{White}(1963)}]{white1963}
{White}, O.~R. 1963, \apj, 138, 1316

\bibitem[{{Wilson} \& {Vainu Bappu}(1957)}]{wilson+bappu1957}
{Wilson}, O.~C. \& {Vainu Bappu}, M.~K. 1957, \apj, 125, 661

\bibitem[{{Zaqarashvili} {et~al.}(2010){Zaqarashvili}, {Kukhianidze}, \&
  {Khodachenko}}]{zaqarashvili+etal2010}
{Zaqarashvili}, T.~V., {Kukhianidze}, V., \& {Khodachenko}, M. 2010, \mnras,
  404, L74

\bibitem[{{Zaqarashvili} \& {Skhirtladze}(2008)}]{zaqarashvili+etal2008}
{Zaqarashvili}, T.~V. \& {Skhirtladze}, N. 2008, \apjl, 683, L91

\bibitem[{{Zirker}(1962)}]{zirker1962}
{Zirker}, J.~B. 1962, \apj, 136, 250

\bibitem[{{Zirker}(1968)}]{zirker1968}
{Zirker}, J.~B. 1968, \solphys, 3, 164

\bibitem[{{Zwaan}(1965)}]{zwaan1965}
{Zwaan}, C. 1965, Recherches Astronomiques de l'Observatoire d'Utrecht, 17

\end{thebibliography}
\clearpage
\begin{appendix}
\section{Spatial resolution and residual image motion \label{spat_res_sect}}
Because observations near the solar limb are usually difficult, especially when one tries to maintain the AO real-time correction, we investigate the achieved spatial resolution. Figure \ref{spat_res} shows two estimates of the spatial resolution, from the Fourier power as function of the spatial frequency ({\em left}) and the extension of an isolated brightening in the H$_{\rm 2R}$ map ({\em right}). The spatial sampling provides a maximal resolution of 0\farcs6. The Fourier power levels off to a constant value that is indicative of the noise level between 1$^{\prime\prime}$ and 1.5$^{\prime\prime}$. The shape of the curves for different wavelength windows (OW, H-index, fixed wavelength at 396.822\,nm) or the difference between H$_{\rm 2V}$ and H$_{\rm 2R}$ is fairly similar, with slightly enhanced power for the intensity difference measure. A Gaussian fit to the extension of the isolated brightening shown as inset in the {\em right panel} yielded a FWHM of 1.5$^{\prime\prime}$ and 1.9$^{\prime\prime}$ on cuts in the horizontal and vertical direction, respectively, with the caveat that the feature presumably will also have an intrinsic spatial extent. The spatial resolution should accordingly be about 1.5$^{\prime\prime}$ or slightly better.
\begin{figure}
\centerline{\resizebox{8.8cm}{!}{\includegraphics{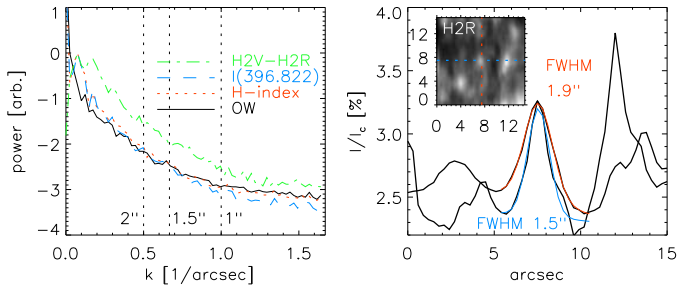}}}
\caption{Spatial resolution. {\em Left}: Fourier power as function of spatial frequency. {\em Right}: vertical and horizontal cuts through the center of the H$_{\rm 2R}$ map shown as inset at the {\em upper left}. Gaussian fits to the central brightening are overplotted with {\em red} and {\em blue} lines, respectively.\label{spat_res}}
\end{figure}
\begin{figure}
\centerline{\resizebox{8.8cm}{!}{\includegraphics{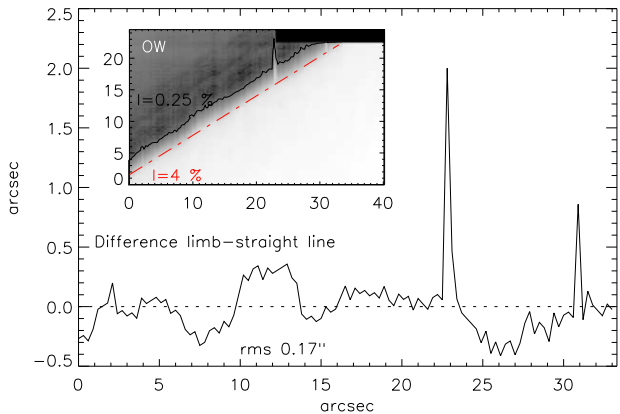}}}
\caption{Position of the limb in the OW map (inset at the {\em upper left} in logarithmic scaling). The {\em inclined dash-dotted line} in the inset denotes a linear fit to the limb position. The {\em black line} denotes an intensity level of 0.25\,\%. The line plot below the inset shows the difference between the {\em dash-dotted} line and the actual limb position in arcsec. \label{limb_pos1}}
\end{figure}
\begin{figure}
\centerline{\resizebox{8.8cm}{!}{\includegraphics{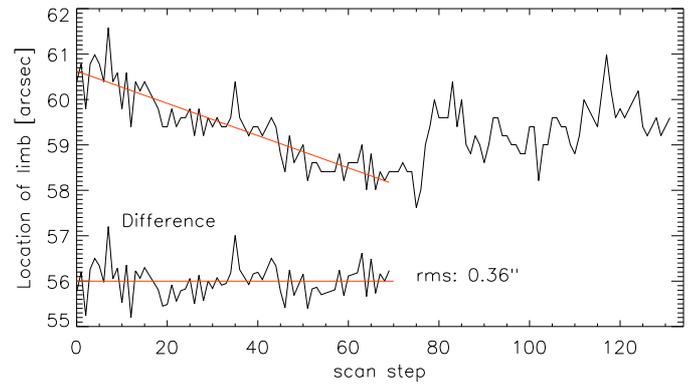}}}
\caption{Location of the limb for one spatial position in the SJ images. The {\em lower part} shows the difference between the {\em red} straight line and the limb position. \label{limb_pos2}}
\end{figure}

To determine the residual image motion caused by the variable seeing and uncompensated image shifts, we measured the location of the ``white-light'' limb in the OW map (inset in Fig.~\ref{limb_pos1}). The limb was defined as the point where the intensity dropped below 4\%. We fitted a straight line to the limb location and then derived the difference to the actual limb position at one scan step and the straight line (Fig.~\ref{limb_pos1}). The rms variation of the difference is about 0\farcs17. Because this is now derived from the spectra after the 13-sec integration, it only provides a temporally averaged value of the image motion. We therefore used the individual slit-jaw (SJ) images for a cross-check. The SJ images are taken at the beginning of each scan step with an exposure time below 20\,ms; they are thus not influenced by any temporal averaging. We removed the image motion caused by the spatial scanning perpendicular to the slit from the SJ images, and then determined the location of the limb in the individual images, defined again as the position where the intensity drops below 4\%. Figure \ref{limb_pos2} shows the distance between the limb and the edge of the SJ image for one randomly chosen image row. In addition to a linear trend that could be caused by a slow drift of the AO lock point or a small tilt angle between the POLIS scan mirror and the slit, the limb position in the SJ images fluctuated with an rms amplitude of about 0\farcs36. The fluctuations of the atmospheric scattering were below a level of 0.25\,\% in the OW map, the map is still fairly homogeneous at this intensity level ({\em black line} in the inset of in Fig.~\ref{limb_pos1}), with the distance to the limb being the dominating factor for the residual light level.
\section{Additional example spectra\label{appa}}
Figure \ref{multi_lobe} shows an example of the multi-lobed profiles found on the disk near a patch of enhanced emission and strong photospheric polarization signal. Figure \ref{fig7_app} shows a series of individual neighboring spectra at a limb distance of 2.5\,Mm. The Doppler shift of the absorption core relative to the emission peaks reverses from blue to red shift over a spatial scale of about 1.6$^{\prime\prime}$. The spectra along the slit (Fig.~\ref{fig17}) show the same patterns as those along cuts perpendicular to the limb: a variation of Doppler shifts both in the horizontal direction from scan step to scan step as well as a variation in single cuts along the vertical axis. Figures \ref{fig22} and \ref{fig23} show the individual profiles of cut no.~2 and the scan step no.~66. In the first case, the distance given at the {\em left} of each profile is measured perpendicular to the limb, in the latter case it only reflects the location along the slit.
\begin{figure}
\centerline{\resizebox{8.8cm}{!}{\includegraphics{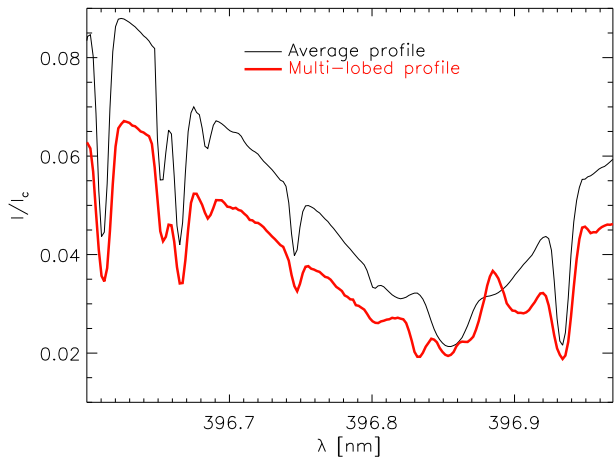}}}
\caption{Ca profile with more than two reversals ({\em thick red}). The {\em thin black line} shows the average profile for comparison.\label{multi_lobe}}
\end{figure}
\begin{figure}
\centerline{\resizebox{8.8cm}{!}{\includegraphics{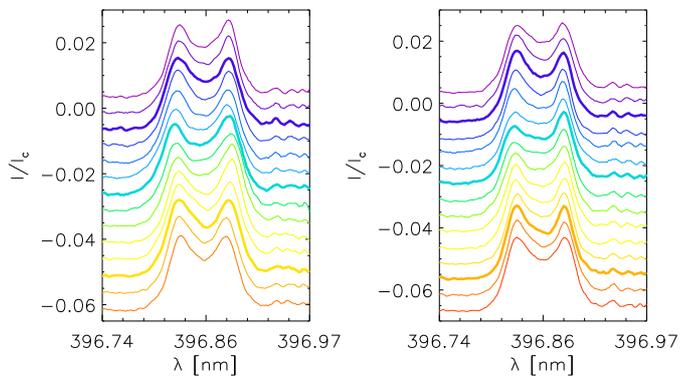}}}
\caption{Individual spectra at a limb distance of 2.5 Mm. {\em Left}: spectra from $x$=4.5$^{\prime\prime}$ to 8.4$^{\prime\prime}$. {\em Thick lines} indicate profiles with large shifts. {\em Right}: same from $x$=9.3$^{\prime\prime}$ to 13.5$^{\prime\prime}$. The position values refer to the spatial scale given in Fig.~\ref{fig6}.\label{fig7_app}}
\end{figure}
\begin{figure*}
\sidecaption
\begin{minipage}{12cm}
\centerline{\resizebox{11.cm}{!}{\includegraphics{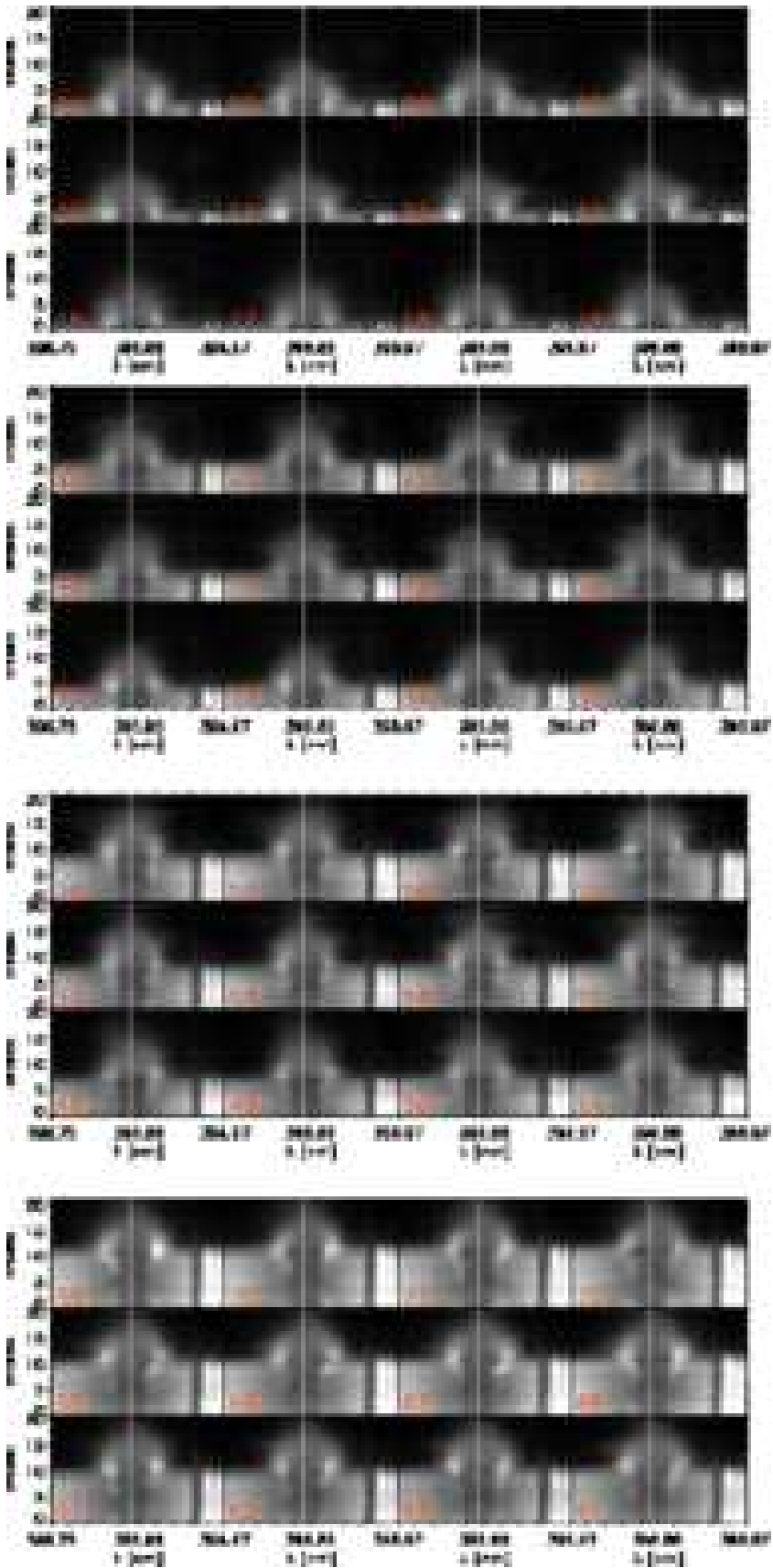}}}
\end{minipage}
\caption{Spectra along the slit around the scan step 21 ({\em top panel}), step 37 ({\em second panel}), step 53 ({\em third panel}), and step 66 ({\em bottom panel}). The scan step of each cut along the slit is given at the {\em lower left} of each panel, increasing from {\em left to right} and from {\em bottom up}. The $y$-axis gives the distance along the cut, the limb is thus located at varying positions. The {\em vertical white line} denotes the rest wavelength of \ion{Ca}{ii}\,H.\label{fig17}}
\end{figure*}
\clearpage
\begin{figure}
\centerline{\resizebox{8.cm}{!}{\includegraphics{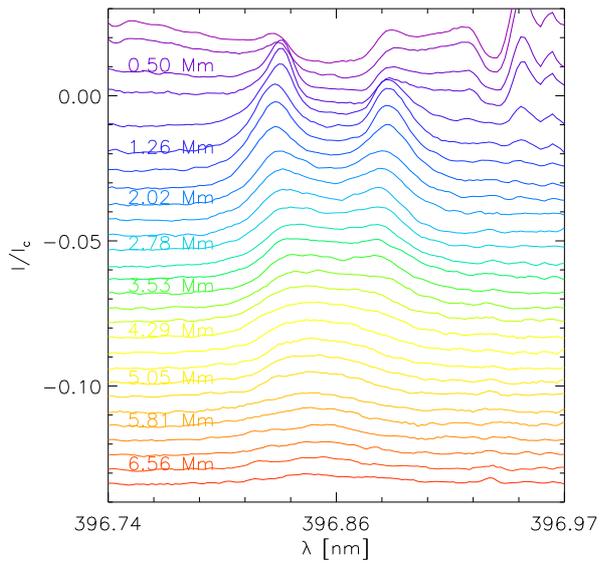}}}
\caption{Spectra of cut no.~2 (see Fig.~\ref{fig2b}) with increasing limb distance. The limb is at the top, each third profile is labeled with its corresponding height.\label{fig22}}
\end{figure}
\begin{figure}
\centerline{\resizebox{8.cm}{!}{\includegraphics{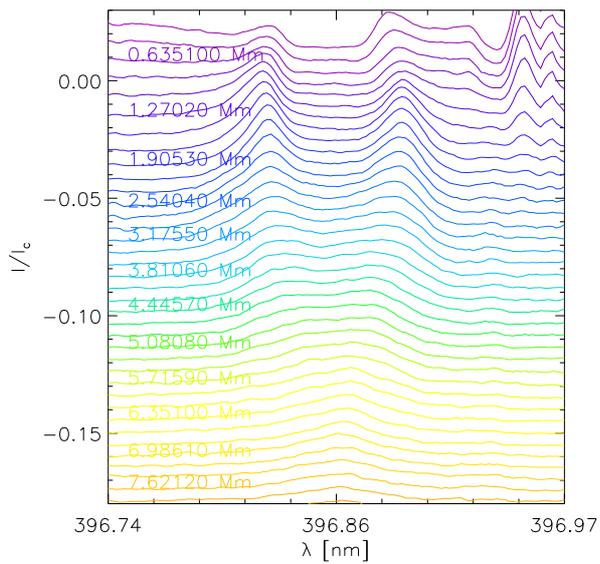}}}
\caption{Spectra of scan step no.~66 (see Fig.~\ref{fig2b}) with increasing position along the slit. The limb is at the top, each third profile is labeled with its corresponding height.\label{fig23}}
\end{figure}
\end{appendix}
\end{document}